\documentclass[twocolumn,tighten]{aastex61}
\usepackage[]{txfonts}
\usepackage{natbib, twoopt}
\usepackage{float}
\usepackage{graphicx}
\usepackage{epstopdf}
\usepackage{enumitem}
\usepackage[caption=false]{subfig}
\usepackage{color}\definecolor{AliceBlue}{rgb}{0.94,0.97,1.00}
\definecolor{AntiqueWhite1}{rgb}{1.00,0.94,0.86}
\definecolor{AntiqueWhite2}{rgb}{0.93,0.87,0.80}
\definecolor{AntiqueWhite3}{rgb}{0.80,0.75,0.69}
\definecolor{AntiqueWhite4}{rgb}{0.55,0.51,0.47}
\definecolor{AntiqueWhite}{rgb}{0.98,0.92,0.84}
\definecolor{BlanchedAlmond}{rgb}{1.00,0.92,0.80}
\definecolor{BlueViolet}{rgb}{0.54,0.17,0.89}
\definecolor{CadetBlue1}{rgb}{0.60,0.96,1.00}
\definecolor{CadetBlue2}{rgb}{0.56,0.90,0.93}
\definecolor{CadetBlue3}{rgb}{0.48,0.77,0.80}
\definecolor{CadetBlue4}{rgb}{0.33,0.53,0.55}
\definecolor{CadetBlue}{rgb}{0.37,0.62,0.63}
\definecolor{CornflowerBlue}{rgb}{0.39,0.58,0.93}
\definecolor{DarkBlue}{rgb}{0.00,0.00,0.55}
\definecolor{DarkCyan}{rgb}{0.00,0.55,0.55}
\definecolor{DarkGoldenrod1}{rgb}{1.00,0.73,0.06}
\definecolor{DarkGoldenrod2}{rgb}{0.93,0.68,0.05}
\definecolor{DarkGoldenrod3}{rgb}{0.80,0.58,0.05}
\definecolor{DarkGoldenrod4}{rgb}{0.55,0.40,0.03}
\definecolor{DarkGoldenrod}{rgb}{0.72,0.53,0.04}
\definecolor{DarkGray}{rgb}{0.66,0.66,0.66}
\definecolor{DarkGreen}{rgb}{0.00,0.39,0.00}
\definecolor{DarkGrey}{rgb}{0.66,0.66,0.66}
\definecolor{DarkKhaki}{rgb}{0.74,0.72,0.42}
\definecolor{DarkMagenta}{rgb}{0.55,0.00,0.55}
\definecolor{DarkOliveGreen1}{rgb}{0.79,1.00,0.44}
\definecolor{DarkOliveGreen2}{rgb}{0.74,0.93,0.41}
\definecolor{DarkOliveGreen3}{rgb}{0.64,0.80,0.35}
\definecolor{DarkOliveGreen4}{rgb}{0.43,0.55,0.24}
\definecolor{DarkOliveGreen}{rgb}{0.33,0.42,0.18}
\definecolor{DarkOrange1}{rgb}{1.00,0.50,0.00}
\definecolor{DarkOrange2}{rgb}{0.93,0.46,0.00}
\definecolor{DarkOrange3}{rgb}{0.80,0.40,0.00}
\definecolor{DarkOrange4}{rgb}{0.55,0.27,0.00}
\definecolor{DarkOrange}{rgb}{1.00,0.55,0.00}
\definecolor{DarkOrchid1}{rgb}{0.75,0.24,1.00}
\definecolor{DarkOrchid2}{rgb}{0.70,0.23,0.93}
\definecolor{DarkOrchid3}{rgb}{0.60,0.20,0.80}
\definecolor{DarkOrchid4}{rgb}{0.41,0.13,0.55}
\definecolor{DarkOrchid}{rgb}{0.60,0.20,0.80}
\definecolor{DarkRed}{rgb}{0.55,0.00,0.00}
\definecolor{DarkSalmon}{rgb}{0.91,0.59,0.48}
\definecolor{DarkSeaGreen1}{rgb}{0.76,1.00,0.76}
\definecolor{DarkSeaGreen2}{rgb}{0.71,0.93,0.71}
\definecolor{DarkSeaGreen3}{rgb}{0.61,0.80,0.61}
\definecolor{DarkSeaGreen4}{rgb}{0.41,0.55,0.41}
\definecolor{DarkSeaGreen}{rgb}{0.56,0.74,0.56}
\definecolor{DarkSlateBlue}{rgb}{0.28,0.24,0.55}
\definecolor{DarkSlateGray1}{rgb}{0.59,1.00,1.00}
\definecolor{DarkSlateGray2}{rgb}{0.55,0.93,0.93}
\definecolor{DarkSlateGray3}{rgb}{0.47,0.80,0.80}
\definecolor{DarkSlateGray4}{rgb}{0.32,0.55,0.55}
\definecolor{DarkSlateGray}{rgb}{0.18,0.31,0.31}
\definecolor{DarkSlateGrey}{rgb}{0.18,0.31,0.31}
\definecolor{DarkTurquoise}{rgb}{0.00,0.81,0.82}
\definecolor{DarkViolet}{rgb}{0.58,0.00,0.83}
\definecolor{DeepPink1}{rgb}{1.00,0.08,0.58}
\definecolor{DeepPink2}{rgb}{0.93,0.07,0.54}
\definecolor{DeepPink3}{rgb}{0.80,0.06,0.46}
\definecolor{DeepPink4}{rgb}{0.55,0.04,0.31}
\definecolor{DeepPink}{rgb}{1.00,0.08,0.58}
\definecolor{DeepSkyBlue1}{rgb}{0.00,0.75,1.00}
\definecolor{DeepSkyBlue2}{rgb}{0.00,0.70,0.93}
\definecolor{DeepSkyBlue3}{rgb}{0.00,0.60,0.80}
\definecolor{DeepSkyBlue4}{rgb}{0.00,0.41,0.55}
\definecolor{DeepSkyBlue}{rgb}{0.00,0.75,1.00}
\definecolor{DimGray}{rgb}{0.41,0.41,0.41}
\definecolor{DimGrey}{rgb}{0.41,0.41,0.41}
\definecolor{DodgerBlue1}{rgb}{0.12,0.56,1.00}
\definecolor{DodgerBlue2}{rgb}{0.11,0.53,0.93}
\definecolor{DodgerBlue3}{rgb}{0.09,0.45,0.80}
\definecolor{DodgerBlue4}{rgb}{0.06,0.31,0.55}
\definecolor{DodgerBlue}{rgb}{0.12,0.56,1.00}
\definecolor{FloralWhite}{rgb}{1.00,0.98,0.94}
\definecolor{ForestGreen}{rgb}{0.13,0.55,0.13}
\definecolor{GhostWhite}{rgb}{0.97,0.97,1.00}
\definecolor{GreenYellow}{rgb}{0.68,1.00,0.18}
\definecolor{HotPink1}{rgb}{1.00,0.43,0.71}
\definecolor{HotPink2}{rgb}{0.93,0.42,0.65}
\definecolor{HotPink3}{rgb}{0.80,0.38,0.56}
\definecolor{HotPink4}{rgb}{0.55,0.23,0.38}
\definecolor{HotPink}{rgb}{1.00,0.41,0.71}
\definecolor{IndianRed1}{rgb}{1.00,0.42,0.42}
\definecolor{IndianRed2}{rgb}{0.93,0.39,0.39}
\definecolor{IndianRed3}{rgb}{0.80,0.33,0.33}
\definecolor{IndianRed4}{rgb}{0.55,0.23,0.23}
\definecolor{IndianRed}{rgb}{0.80,0.36,0.36}
\definecolor{LavenderBlush1}{rgb}{1.00,0.94,0.96}
\definecolor{LavenderBlush2}{rgb}{0.93,0.88,0.90}
\definecolor{LavenderBlush3}{rgb}{0.80,0.76,0.77}
\definecolor{LavenderBlush4}{rgb}{0.55,0.51,0.53}
\definecolor{LavenderBlush}{rgb}{1.00,0.94,0.96}
\definecolor{LawnGreen}{rgb}{0.49,0.99,0.00}
\definecolor{LemonChiffon1}{rgb}{1.00,0.98,0.80}
\definecolor{LemonChiffon2}{rgb}{0.93,0.91,0.75}
\definecolor{LemonChiffon3}{rgb}{0.80,0.79,0.65}
\definecolor{LemonChiffon4}{rgb}{0.55,0.54,0.44}
\definecolor{LemonChiffon}{rgb}{1.00,0.98,0.80}
\definecolor{LightBlue1}{rgb}{0.75,0.94,1.00}
\definecolor{LightBlue2}{rgb}{0.70,0.87,0.93}
\definecolor{LightBlue3}{rgb}{0.60,0.75,0.80}
\definecolor{LightBlue4}{rgb}{0.41,0.51,0.55}
\definecolor{LightBlue}{rgb}{0.68,0.85,0.90}
\definecolor{LightCoral}{rgb}{0.94,0.50,0.50}
\definecolor{LightCyan1}{rgb}{0.88,1.00,1.00}
\definecolor{LightCyan2}{rgb}{0.82,0.93,0.93}
\definecolor{LightCyan3}{rgb}{0.71,0.80,0.80}
\definecolor{LightCyan4}{rgb}{0.48,0.55,0.55}
\definecolor{LightCyan}{rgb}{0.88,1.00,1.00}
\definecolor{LightGoldenrod1}{rgb}{1.00,0.93,0.55}
\definecolor{LightGoldenrod2}{rgb}{0.93,0.86,0.51}
\definecolor{LightGoldenrod3}{rgb}{0.80,0.75,0.44}
\definecolor{LightGoldenrod4}{rgb}{0.55,0.51,0.30}
\definecolor{LightGoldenrodYellow}{rgb}{0.98,0.98,0.82}
\definecolor{LightGoldenrod}{rgb}{0.93,0.87,0.51}
\definecolor{LightGray}{rgb}{0.83,0.83,0.83}
\definecolor{LightGreen}{rgb}{0.56,0.93,0.56}
\definecolor{LightGrey}{rgb}{0.83,0.83,0.83}
\definecolor{LightPink1}{rgb}{1.00,0.68,0.73}
\definecolor{LightPink2}{rgb}{0.93,0.64,0.68}
\definecolor{LightPink3}{rgb}{0.80,0.55,0.58}
\definecolor{LightPink4}{rgb}{0.55,0.37,0.40}
\definecolor{LightPink}{rgb}{1.00,0.71,0.76}
\definecolor{LightSalmon1}{rgb}{1.00,0.63,0.48}
\definecolor{LightSalmon2}{rgb}{0.93,0.58,0.45}
\definecolor{LightSalmon3}{rgb}{0.80,0.51,0.38}
\definecolor{LightSalmon4}{rgb}{0.55,0.34,0.26}
\definecolor{LightSalmon}{rgb}{1.00,0.63,0.48}
\definecolor{LightSeaGreen}{rgb}{0.13,0.70,0.67}
\definecolor{LightSkyBlue1}{rgb}{0.69,0.89,1.00}
\definecolor{LightSkyBlue2}{rgb}{0.64,0.83,0.93}
\definecolor{LightSkyBlue3}{rgb}{0.55,0.71,0.80}
\definecolor{LightSkyBlue4}{rgb}{0.38,0.48,0.55}
\definecolor{LightSkyBlue}{rgb}{0.53,0.81,0.98}
\definecolor{LightSlateBlue}{rgb}{0.52,0.44,1.00}
\definecolor{LightSlateGray}{rgb}{0.47,0.53,0.60}
\definecolor{LightSlateGrey}{rgb}{0.47,0.53,0.60}
\definecolor{LightSteelBlue1}{rgb}{0.79,0.88,1.00}
\definecolor{LightSteelBlue2}{rgb}{0.74,0.82,0.93}
\definecolor{LightSteelBlue3}{rgb}{0.64,0.71,0.80}
\definecolor{LightSteelBlue4}{rgb}{0.43,0.48,0.55}
\definecolor{LightSteelBlue}{rgb}{0.69,0.77,0.87}
\definecolor{LightYellow1}{rgb}{1.00,1.00,0.88}
\definecolor{LightYellow2}{rgb}{0.93,0.93,0.82}
\definecolor{LightYellow3}{rgb}{0.80,0.80,0.71}
\definecolor{LightYellow4}{rgb}{0.55,0.55,0.48}
\definecolor{LightYellow}{rgb}{1.00,1.00,0.88}
\definecolor{LimeGreen}{rgb}{0.20,0.80,0.20}
\definecolor{MediumAquamarine}{rgb}{0.40,0.80,0.67}
\definecolor{MediumBlue}{rgb}{0.00,0.00,0.80}
\definecolor{MediumOrchid1}{rgb}{0.88,0.40,1.00}
\definecolor{MediumOrchid2}{rgb}{0.82,0.37,0.93}
\definecolor{MediumOrchid3}{rgb}{0.71,0.32,0.80}
\definecolor{MediumOrchid4}{rgb}{0.48,0.22,0.55}
\definecolor{MediumOrchid}{rgb}{0.73,0.33,0.83}
\definecolor{MediumPurple1}{rgb}{0.67,0.51,1.00}
\definecolor{MediumPurple2}{rgb}{0.62,0.47,0.93}
\definecolor{MediumPurple3}{rgb}{0.54,0.41,0.80}
\definecolor{MediumPurple4}{rgb}{0.36,0.28,0.55}
\definecolor{MediumPurple}{rgb}{0.58,0.44,0.86}
\definecolor{MediumSeaGreen}{rgb}{0.24,0.70,0.44}
\definecolor{MediumSlateBlue}{rgb}{0.48,0.41,0.93}
\definecolor{MediumSpringGreen}{rgb}{0.00,0.98,0.60}
\definecolor{MediumTurquoise}{rgb}{0.28,0.82,0.80}
\definecolor{MediumVioletRed}{rgb}{0.78,0.08,0.52}
\definecolor{MidnightBlue}{rgb}{0.10,0.10,0.44}
\definecolor{MintCream}{rgb}{0.96,1.00,0.98}
\definecolor{MistyRose1}{rgb}{1.00,0.89,0.88}
\definecolor{MistyRose2}{rgb}{0.93,0.84,0.82}
\definecolor{MistyRose3}{rgb}{0.80,0.72,0.71}
\definecolor{MistyRose4}{rgb}{0.55,0.49,0.48}
\definecolor{MistyRose}{rgb}{1.00,0.89,0.88}
\definecolor{NavajoWhite1}{rgb}{1.00,0.87,0.68}
\definecolor{NavajoWhite2}{rgb}{0.93,0.81,0.63}
\definecolor{NavajoWhite3}{rgb}{0.80,0.70,0.55}
\definecolor{NavajoWhite4}{rgb}{0.55,0.47,0.37}
\definecolor{NavajoWhite}{rgb}{1.00,0.87,0.68}
\definecolor{NavyBlue}{rgb}{0.00,0.00,0.50}
\definecolor{OldLace}{rgb}{0.99,0.96,0.90}
\definecolor{OliveDrab1}{rgb}{0.75,1.00,0.24}
\definecolor{OliveDrab2}{rgb}{0.70,0.93,0.23}
\definecolor{OliveDrab3}{rgb}{0.60,0.80,0.20}
\definecolor{OliveDrab4}{rgb}{0.41,0.55,0.13}
\definecolor{OliveDrab}{rgb}{0.42,0.56,0.14}
\definecolor{OrangeRed1}{rgb}{1.00,0.27,0.00}
\definecolor{OrangeRed2}{rgb}{0.93,0.25,0.00}
\definecolor{OrangeRed3}{rgb}{0.80,0.22,0.00}
\definecolor{OrangeRed4}{rgb}{0.55,0.15,0.00}
\definecolor{OrangeRed}{rgb}{1.00,0.27,0.00}
\definecolor{PaleGoldenrod}{rgb}{0.93,0.91,0.67}
\definecolor{PaleGreen1}{rgb}{0.60,1.00,0.60}
\definecolor{PaleGreen2}{rgb}{0.56,0.93,0.56}
\definecolor{PaleGreen3}{rgb}{0.49,0.80,0.49}
\definecolor{PaleGreen4}{rgb}{0.33,0.55,0.33}
\definecolor{PaleGreen}{rgb}{0.60,0.98,0.60}
\definecolor{PaleTurquoise1}{rgb}{0.73,1.00,1.00}
\definecolor{PaleTurquoise2}{rgb}{0.68,0.93,0.93}
\definecolor{PaleTurquoise3}{rgb}{0.59,0.80,0.80}
\definecolor{PaleTurquoise4}{rgb}{0.40,0.55,0.55}
\definecolor{PaleTurquoise}{rgb}{0.69,0.93,0.93}
\definecolor{PaleVioletRed1}{rgb}{1.00,0.51,0.67}
\definecolor{PaleVioletRed2}{rgb}{0.93,0.47,0.62}
\definecolor{PaleVioletRed3}{rgb}{0.80,0.41,0.54}
\definecolor{PaleVioletRed4}{rgb}{0.55,0.28,0.36}
\definecolor{PaleVioletRed}{rgb}{0.86,0.44,0.58}
\definecolor{PapayaWhip}{rgb}{1.00,0.94,0.84}
\definecolor{PeachPuff1}{rgb}{1.00,0.85,0.73}
\definecolor{PeachPuff2}{rgb}{0.93,0.80,0.68}
\definecolor{PeachPuff3}{rgb}{0.80,0.69,0.58}
\definecolor{PeachPuff4}{rgb}{0.55,0.47,0.40}
\definecolor{PeachPuff}{rgb}{1.00,0.85,0.73}
\definecolor{PowderBlue}{rgb}{0.69,0.88,0.90}
\definecolor{RosyBrown1}{rgb}{1.00,0.76,0.76}
\definecolor{RosyBrown2}{rgb}{0.93,0.71,0.71}
\definecolor{RosyBrown3}{rgb}{0.80,0.61,0.61}
\definecolor{RosyBrown4}{rgb}{0.55,0.41,0.41}
\definecolor{RosyBrown}{rgb}{0.74,0.56,0.56}
\definecolor{RoyalBlue1}{rgb}{0.28,0.46,1.00}
\definecolor{RoyalBlue2}{rgb}{0.26,0.43,0.93}
\definecolor{RoyalBlue3}{rgb}{0.23,0.37,0.80}
\definecolor{RoyalBlue4}{rgb}{0.15,0.25,0.55}
\definecolor{RoyalBlue}{rgb}{0.25,0.41,0.88}
\definecolor{SaddleBrown}{rgb}{0.55,0.27,0.07}
\definecolor{SandyBrown}{rgb}{0.96,0.64,0.38}
\definecolor{SeaGreen1}{rgb}{0.33,1.00,0.62}
\definecolor{SeaGreen2}{rgb}{0.31,0.93,0.58}
\definecolor{SeaGreen3}{rgb}{0.26,0.80,0.50}
\definecolor{SeaGreen4}{rgb}{0.18,0.55,0.34}
\definecolor{SeaGreen}{rgb}{0.18,0.55,0.34}
\definecolor{SkyBlue1}{rgb}{0.53,0.81,1.00}
\definecolor{SkyBlue2}{rgb}{0.49,0.75,0.93}
\definecolor{SkyBlue3}{rgb}{0.42,0.65,0.80}
\definecolor{SkyBlue4}{rgb}{0.29,0.44,0.55}
\definecolor{SkyBlue}{rgb}{0.53,0.81,0.92}
\definecolor{SlateBlue1}{rgb}{0.51,0.44,1.00}
\definecolor{SlateBlue2}{rgb}{0.48,0.40,0.93}
\definecolor{SlateBlue3}{rgb}{0.41,0.35,0.80}
\definecolor{SlateBlue4}{rgb}{0.28,0.24,0.55}
\definecolor{SlateBlue}{rgb}{0.42,0.35,0.80}
\definecolor{SlateGray1}{rgb}{0.78,0.89,1.00}
\definecolor{SlateGray2}{rgb}{0.73,0.83,0.93}
\definecolor{SlateGray3}{rgb}{0.62,0.71,0.80}
\definecolor{SlateGray4}{rgb}{0.42,0.48,0.55}
\definecolor{SlateGray}{rgb}{0.44,0.50,0.56}
\definecolor{SlateGrey}{rgb}{0.44,0.50,0.56}
\definecolor{SpringGreen1}{rgb}{0.00,1.00,0.50}
\definecolor{SpringGreen2}{rgb}{0.00,0.93,0.46}
\definecolor{SpringGreen3}{rgb}{0.00,0.80,0.40}
\definecolor{SpringGreen4}{rgb}{0.00,0.55,0.27}
\definecolor{SpringGreen}{rgb}{0.00,1.00,0.50}
\definecolor{SteelBlue1}{rgb}{0.39,0.72,1.00}
\definecolor{SteelBlue2}{rgb}{0.36,0.67,0.93}
\definecolor{SteelBlue3}{rgb}{0.31,0.58,0.80}
\definecolor{SteelBlue4}{rgb}{0.21,0.39,0.55}
\definecolor{SteelBlue}{rgb}{0.27,0.51,0.71}
\definecolor{VioletRed1}{rgb}{1.00,0.24,0.59}
\definecolor{VioletRed2}{rgb}{0.93,0.23,0.55}
\definecolor{VioletRed3}{rgb}{0.80,0.20,0.47}
\definecolor{VioletRed4}{rgb}{0.55,0.13,0.32}
\definecolor{VioletRed}{rgb}{0.82,0.13,0.56}
\definecolor{WhiteSmoke}{rgb}{0.96,0.96,0.96}
\definecolor{YellowGreen}{rgb}{0.60,0.80,0.20}
\definecolor{aliceblue}{rgb}{0.94,0.97,1.00}
\definecolor{antiquewhite}{rgb}{0.98,0.92,0.84}
\definecolor{aquamarine1}{rgb}{0.50,1.00,0.83}
\definecolor{aquamarine2}{rgb}{0.46,0.93,0.78}
\definecolor{aquamarine3}{rgb}{0.40,0.80,0.67}
\definecolor{aquamarine4}{rgb}{0.27,0.55,0.45}
\definecolor{aquamarine}{rgb}{0.50,1.00,0.83}
\definecolor{azure1}{rgb}{0.94,1.00,1.00}
\definecolor{azure2}{rgb}{0.88,0.93,0.93}
\definecolor{azure3}{rgb}{0.76,0.80,0.80}
\definecolor{azure4}{rgb}{0.51,0.55,0.55}
\definecolor{azure}{rgb}{0.94,1.00,1.00}
\definecolor{beige}{rgb}{0.96,0.96,0.86}
\definecolor{bisque1}{rgb}{1.00,0.89,0.77}
\definecolor{bisque2}{rgb}{0.93,0.84,0.72}
\definecolor{bisque3}{rgb}{0.80,0.72,0.62}
\definecolor{bisque4}{rgb}{0.55,0.49,0.42}
\definecolor{bisque}{rgb}{1.00,0.89,0.77}
\definecolor{black}{rgb}{0.00,0.00,0.00}
\definecolor{blanchedalmond}{rgb}{1.00,0.92,0.80}
\definecolor{blue1}{rgb}{0.00,0.00,1.00}
\definecolor{blue2}{rgb}{0.00,0.00,0.93}
\definecolor{blue3}{rgb}{0.00,0.00,0.80}
\definecolor{blue4}{rgb}{0.00,0.00,0.55}
\definecolor{blueviolet}{rgb}{0.54,0.17,0.89}
\definecolor{blue}{rgb}{0.00,0.00,1.00}
\definecolor{brown1}{rgb}{1.00,0.25,0.25}
\definecolor{brown2}{rgb}{0.93,0.23,0.23}
\definecolor{brown3}{rgb}{0.80,0.20,0.20}
\definecolor{brown4}{rgb}{0.55,0.14,0.14}
\definecolor{brown}{rgb}{0.65,0.16,0.16}
\definecolor{burlywood1}{rgb}{1.00,0.83,0.61}
\definecolor{burlywood2}{rgb}{0.93,0.77,0.57}
\definecolor{burlywood3}{rgb}{0.80,0.67,0.49}
\definecolor{burlywood4}{rgb}{0.55,0.45,0.33}
\definecolor{burlywood}{rgb}{0.87,0.72,0.53}
\definecolor{cadetblue}{rgb}{0.37,0.62,0.63}
\definecolor{chartreuse1}{rgb}{0.50,1.00,0.00}
\definecolor{chartreuse2}{rgb}{0.46,0.93,0.00}
\definecolor{chartreuse3}{rgb}{0.40,0.80,0.00}
\definecolor{chartreuse4}{rgb}{0.27,0.55,0.00}
\definecolor{chartreuse}{rgb}{0.50,1.00,0.00}
\definecolor{chocolate1}{rgb}{1.00,0.50,0.14}
\definecolor{chocolate2}{rgb}{0.93,0.46,0.13}
\definecolor{chocolate3}{rgb}{0.80,0.40,0.11}
\definecolor{chocolate4}{rgb}{0.55,0.27,0.07}
\definecolor{chocolate}{rgb}{0.82,0.41,0.12}
\definecolor{coral1}{rgb}{1.00,0.45,0.34}
\definecolor{coral2}{rgb}{0.93,0.42,0.31}
\definecolor{coral3}{rgb}{0.80,0.36,0.27}
\definecolor{coral4}{rgb}{0.55,0.24,0.18}
\definecolor{coral}{rgb}{1.00,0.50,0.31}
\definecolor{cornflowerblue}{rgb}{0.39,0.58,0.93}
\definecolor{cornsilk1}{rgb}{1.00,0.97,0.86}
\definecolor{cornsilk2}{rgb}{0.93,0.91,0.80}
\definecolor{cornsilk3}{rgb}{0.80,0.78,0.69}
\definecolor{cornsilk4}{rgb}{0.55,0.53,0.47}
\definecolor{cornsilk}{rgb}{1.00,0.97,0.86}
\definecolor{cyan1}{rgb}{0.00,1.00,1.00}
\definecolor{cyan2}{rgb}{0.00,0.93,0.93}
\definecolor{cyan3}{rgb}{0.00,0.80,0.80}
\definecolor{cyan4}{rgb}{0.00,0.55,0.55}
\definecolor{cyan}{rgb}{0.00,1.00,1.00}
\definecolor{darkblue}{rgb}{0.00,0.00,0.55}
\definecolor{darkcyan}{rgb}{0.00,0.55,0.55}
\definecolor{darkgoldenrod}{rgb}{0.72,0.53,0.04}
\definecolor{darkgray}{rgb}{0.66,0.66,0.66}
\definecolor{darkgreen}{rgb}{0.00,0.39,0.00}
\definecolor{darkgrey}{rgb}{0.66,0.66,0.66}
\definecolor{darkkhaki}{rgb}{0.74,0.72,0.42}
\definecolor{darkmagenta}{rgb}{0.55,0.00,0.55}
\definecolor{darkolive}{rgb}{0.33,0.42,0.18}
\definecolor{darkorange}{rgb}{1.00,0.55,0.00}
\definecolor{darkorchid}{rgb}{0.60,0.20,0.80}
\definecolor{darkred}{rgb}{0.55,0.00,0.00}
\definecolor{darksalmon}{rgb}{0.91,0.59,0.48}
\definecolor{darksea}{rgb}{0.56,0.74,0.56}
\definecolor{darkslate}{rgb}{0.18,0.31,0.31}
\definecolor{darkslate}{rgb}{0.18,0.31,0.31}
\definecolor{darkslate}{rgb}{0.28,0.24,0.55}
\definecolor{darkturquoise}{rgb}{0.00,0.81,0.82}
\definecolor{darkviolet}{rgb}{0.58,0.00,0.83}
\definecolor{deeppink}{rgb}{1.00,0.08,0.58}
\definecolor{deepsky}{rgb}{0.00,0.75,1.00}
\definecolor{dimgray}{rgb}{0.41,0.41,0.41}
\definecolor{dimgrey}{rgb}{0.41,0.41,0.41}
\definecolor{dodgerblue}{rgb}{0.12,0.56,1.00}
\definecolor{firebrick1}{rgb}{1.00,0.19,0.19}
\definecolor{firebrick2}{rgb}{0.93,0.17,0.17}
\definecolor{firebrick3}{rgb}{0.80,0.15,0.15}
\definecolor{firebrick4}{rgb}{0.55,0.10,0.10}
\definecolor{firebrick}{rgb}{0.70,0.13,0.13}
\definecolor{floralwhite}{rgb}{1.00,0.98,0.94}
\definecolor{forestgreen}{rgb}{0.13,0.55,0.13}
\definecolor{gainsboro}{rgb}{0.86,0.86,0.86}
\definecolor{ghostwhite}{rgb}{0.97,0.97,1.00}
\definecolor{gold1}{rgb}{1.00,0.84,0.00}
\definecolor{gold2}{rgb}{0.93,0.79,0.00}
\definecolor{gold3}{rgb}{0.80,0.68,0.00}
\definecolor{gold4}{rgb}{0.55,0.46,0.00}
\definecolor{goldenrod1}{rgb}{1.00,0.76,0.15}
\definecolor{goldenrod2}{rgb}{0.93,0.71,0.13}
\definecolor{goldenrod3}{rgb}{0.80,0.61,0.11}
\definecolor{goldenrod4}{rgb}{0.55,0.41,0.08}
\definecolor{goldenrod}{rgb}{0.85,0.65,0.13}
\definecolor{gold}{rgb}{1.00,0.84,0.00}
\definecolor{gray0}{rgb}{0.00,0.00,0.00}
\definecolor{gray100}{rgb}{1.00,1.00,1.00}
\definecolor{gray10}{rgb}{0.10,0.10,0.10}
\definecolor{gray11}{rgb}{0.11,0.11,0.11}
\definecolor{gray12}{rgb}{0.12,0.12,0.12}
\definecolor{gray13}{rgb}{0.13,0.13,0.13}
\definecolor{gray14}{rgb}{0.14,0.14,0.14}
\definecolor{gray15}{rgb}{0.15,0.15,0.15}
\definecolor{gray16}{rgb}{0.16,0.16,0.16}
\definecolor{gray17}{rgb}{0.17,0.17,0.17}
\definecolor{gray18}{rgb}{0.18,0.18,0.18}
\definecolor{gray19}{rgb}{0.19,0.19,0.19}
\definecolor{gray1}{rgb}{0.01,0.01,0.01}
\definecolor{gray20}{rgb}{0.20,0.20,0.20}
\definecolor{gray21}{rgb}{0.21,0.21,0.21}
\definecolor{gray22}{rgb}{0.22,0.22,0.22}
\definecolor{gray23}{rgb}{0.23,0.23,0.23}
\definecolor{gray24}{rgb}{0.24,0.24,0.24}
\definecolor{gray25}{rgb}{0.25,0.25,0.25}
\definecolor{gray26}{rgb}{0.26,0.26,0.26}
\definecolor{gray27}{rgb}{0.27,0.27,0.27}
\definecolor{gray28}{rgb}{0.28,0.28,0.28}
\definecolor{gray29}{rgb}{0.29,0.29,0.29}
\definecolor{gray2}{rgb}{0.02,0.02,0.02}
\definecolor{gray30}{rgb}{0.30,0.30,0.30}
\definecolor{gray31}{rgb}{0.31,0.31,0.31}
\definecolor{gray32}{rgb}{0.32,0.32,0.32}
\definecolor{gray33}{rgb}{0.33,0.33,0.33}
\definecolor{gray34}{rgb}{0.34,0.34,0.34}
\definecolor{gray35}{rgb}{0.35,0.35,0.35}
\definecolor{gray36}{rgb}{0.36,0.36,0.36}
\definecolor{gray37}{rgb}{0.37,0.37,0.37}
\definecolor{gray38}{rgb}{0.38,0.38,0.38}
\definecolor{gray39}{rgb}{0.39,0.39,0.39}
\definecolor{gray3}{rgb}{0.03,0.03,0.03}
\definecolor{gray40}{rgb}{0.40,0.40,0.40}
\definecolor{gray41}{rgb}{0.41,0.41,0.41}
\definecolor{gray42}{rgb}{0.42,0.42,0.42}
\definecolor{gray43}{rgb}{0.43,0.43,0.43}
\definecolor{gray44}{rgb}{0.44,0.44,0.44}
\definecolor{gray45}{rgb}{0.45,0.45,0.45}
\definecolor{gray46}{rgb}{0.46,0.46,0.46}
\definecolor{gray47}{rgb}{0.47,0.47,0.47}
\definecolor{gray48}{rgb}{0.48,0.48,0.48}
\definecolor{gray49}{rgb}{0.49,0.49,0.49}
\definecolor{gray4}{rgb}{0.04,0.04,0.04}
\definecolor{gray50}{rgb}{0.50,0.50,0.50}
\definecolor{gray51}{rgb}{0.51,0.51,0.51}
\definecolor{gray52}{rgb}{0.52,0.52,0.52}
\definecolor{gray53}{rgb}{0.53,0.53,0.53}
\definecolor{gray54}{rgb}{0.54,0.54,0.54}
\definecolor{gray55}{rgb}{0.55,0.55,0.55}
\definecolor{gray56}{rgb}{0.56,0.56,0.56}
\definecolor{gray57}{rgb}{0.57,0.57,0.57}
\definecolor{gray58}{rgb}{0.58,0.58,0.58}
\definecolor{gray59}{rgb}{0.59,0.59,0.59}
\definecolor{gray5}{rgb}{0.05,0.05,0.05}
\definecolor{gray60}{rgb}{0.60,0.60,0.60}
\definecolor{gray61}{rgb}{0.61,0.61,0.61}
\definecolor{gray62}{rgb}{0.62,0.62,0.62}
\definecolor{gray63}{rgb}{0.63,0.63,0.63}
\definecolor{gray64}{rgb}{0.64,0.64,0.64}
\definecolor{gray65}{rgb}{0.65,0.65,0.65}
\definecolor{gray66}{rgb}{0.66,0.66,0.66}
\definecolor{gray67}{rgb}{0.67,0.67,0.67}
\definecolor{gray68}{rgb}{0.68,0.68,0.68}
\definecolor{gray69}{rgb}{0.69,0.69,0.69}
\definecolor{gray6}{rgb}{0.06,0.06,0.06}
\definecolor{gray70}{rgb}{0.70,0.70,0.70}
\definecolor{gray71}{rgb}{0.71,0.71,0.71}
\definecolor{gray72}{rgb}{0.72,0.72,0.72}
\definecolor{gray73}{rgb}{0.73,0.73,0.73}
\definecolor{gray74}{rgb}{0.74,0.74,0.74}
\definecolor{gray75}{rgb}{0.75,0.75,0.75}
\definecolor{gray76}{rgb}{0.76,0.76,0.76}
\definecolor{gray77}{rgb}{0.77,0.77,0.77}
\definecolor{gray78}{rgb}{0.78,0.78,0.78}
\definecolor{gray79}{rgb}{0.79,0.79,0.79}
\definecolor{gray7}{rgb}{0.07,0.07,0.07}
\definecolor{gray80}{rgb}{0.80,0.80,0.80}
\definecolor{gray81}{rgb}{0.81,0.81,0.81}
\definecolor{gray82}{rgb}{0.82,0.82,0.82}
\definecolor{gray83}{rgb}{0.83,0.83,0.83}
\definecolor{gray84}{rgb}{0.84,0.84,0.84}
\definecolor{gray85}{rgb}{0.85,0.85,0.85}
\definecolor{gray86}{rgb}{0.86,0.86,0.86}
\definecolor{gray87}{rgb}{0.87,0.87,0.87}
\definecolor{gray88}{rgb}{0.88,0.88,0.88}
\definecolor{gray89}{rgb}{0.89,0.89,0.89}
\definecolor{gray8}{rgb}{0.08,0.08,0.08}
\definecolor{gray90}{rgb}{0.90,0.90,0.90}
\definecolor{gray91}{rgb}{0.91,0.91,0.91}
\definecolor{gray92}{rgb}{0.92,0.92,0.92}
\definecolor{gray93}{rgb}{0.93,0.93,0.93}
\definecolor{gray94}{rgb}{0.94,0.94,0.94}
\definecolor{gray95}{rgb}{0.95,0.95,0.95}
\definecolor{gray96}{rgb}{0.96,0.96,0.96}
\definecolor{gray97}{rgb}{0.97,0.97,0.97}
\definecolor{gray98}{rgb}{0.98,0.98,0.98}
\definecolor{gray99}{rgb}{0.99,0.99,0.99}
\definecolor{gray9}{rgb}{0.09,0.09,0.09}
\definecolor{gray}{rgb}{0.75,0.75,0.75}
\definecolor{green1}{rgb}{0.00,1.00,0.00}
\definecolor{green2}{rgb}{0.00,0.93,0.00}
\definecolor{green3}{rgb}{0.00,0.80,0.00}
\definecolor{green4}{rgb}{0.00,0.55,0.00}
\definecolor{greenyellow}{rgb}{0.68,1.00,0.18}
\definecolor{green}{rgb}{0.00,1.00,0.00}
\definecolor{grey0}{rgb}{0.00,0.00,0.00}
\definecolor{grey100}{rgb}{1.00,1.00,1.00}
\definecolor{grey10}{rgb}{0.10,0.10,0.10}
\definecolor{grey11}{rgb}{0.11,0.11,0.11}
\definecolor{grey12}{rgb}{0.12,0.12,0.12}
\definecolor{grey13}{rgb}{0.13,0.13,0.13}
\definecolor{grey14}{rgb}{0.14,0.14,0.14}
\definecolor{grey15}{rgb}{0.15,0.15,0.15}
\definecolor{grey16}{rgb}{0.16,0.16,0.16}
\definecolor{grey17}{rgb}{0.17,0.17,0.17}
\definecolor{grey18}{rgb}{0.18,0.18,0.18}
\definecolor{grey19}{rgb}{0.19,0.19,0.19}
\definecolor{grey1}{rgb}{0.01,0.01,0.01}
\definecolor{grey20}{rgb}{0.20,0.20,0.20}
\definecolor{grey21}{rgb}{0.21,0.21,0.21}
\definecolor{grey22}{rgb}{0.22,0.22,0.22}
\definecolor{grey23}{rgb}{0.23,0.23,0.23}
\definecolor{grey24}{rgb}{0.24,0.24,0.24}
\definecolor{grey25}{rgb}{0.25,0.25,0.25}
\definecolor{grey26}{rgb}{0.26,0.26,0.26}
\definecolor{grey27}{rgb}{0.27,0.27,0.27}
\definecolor{grey28}{rgb}{0.28,0.28,0.28}
\definecolor{grey29}{rgb}{0.29,0.29,0.29}
\definecolor{grey2}{rgb}{0.02,0.02,0.02}
\definecolor{grey30}{rgb}{0.30,0.30,0.30}
\definecolor{grey31}{rgb}{0.31,0.31,0.31}
\definecolor{grey32}{rgb}{0.32,0.32,0.32}
\definecolor{grey33}{rgb}{0.33,0.33,0.33}
\definecolor{grey34}{rgb}{0.34,0.34,0.34}
\definecolor{grey35}{rgb}{0.35,0.35,0.35}
\definecolor{grey36}{rgb}{0.36,0.36,0.36}
\definecolor{grey37}{rgb}{0.37,0.37,0.37}
\definecolor{grey38}{rgb}{0.38,0.38,0.38}
\definecolor{grey39}{rgb}{0.39,0.39,0.39}
\definecolor{grey3}{rgb}{0.03,0.03,0.03}
\definecolor{grey40}{rgb}{0.40,0.40,0.40}
\definecolor{grey41}{rgb}{0.41,0.41,0.41}
\definecolor{grey42}{rgb}{0.42,0.42,0.42}
\definecolor{grey43}{rgb}{0.43,0.43,0.43}
\definecolor{grey44}{rgb}{0.44,0.44,0.44}
\definecolor{grey45}{rgb}{0.45,0.45,0.45}
\definecolor{grey46}{rgb}{0.46,0.46,0.46}
\definecolor{grey47}{rgb}{0.47,0.47,0.47}
\definecolor{grey48}{rgb}{0.48,0.48,0.48}
\definecolor{grey49}{rgb}{0.49,0.49,0.49}
\definecolor{grey4}{rgb}{0.04,0.04,0.04}
\definecolor{grey50}{rgb}{0.50,0.50,0.50}
\definecolor{grey51}{rgb}{0.51,0.51,0.51}
\definecolor{grey52}{rgb}{0.52,0.52,0.52}
\definecolor{grey53}{rgb}{0.53,0.53,0.53}
\definecolor{grey54}{rgb}{0.54,0.54,0.54}
\definecolor{grey55}{rgb}{0.55,0.55,0.55}
\definecolor{grey56}{rgb}{0.56,0.56,0.56}
\definecolor{grey57}{rgb}{0.57,0.57,0.57}
\definecolor{grey58}{rgb}{0.58,0.58,0.58}
\definecolor{grey59}{rgb}{0.59,0.59,0.59}
\definecolor{grey5}{rgb}{0.05,0.05,0.05}
\definecolor{grey60}{rgb}{0.60,0.60,0.60}
\definecolor{grey61}{rgb}{0.61,0.61,0.61}
\definecolor{grey62}{rgb}{0.62,0.62,0.62}
\definecolor{grey63}{rgb}{0.63,0.63,0.63}
\definecolor{grey64}{rgb}{0.64,0.64,0.64}
\definecolor{grey65}{rgb}{0.65,0.65,0.65}
\definecolor{grey66}{rgb}{0.66,0.66,0.66}
\definecolor{grey67}{rgb}{0.67,0.67,0.67}
\definecolor{grey68}{rgb}{0.68,0.68,0.68}
\definecolor{grey69}{rgb}{0.69,0.69,0.69}
\definecolor{grey6}{rgb}{0.06,0.06,0.06}
\definecolor{grey70}{rgb}{0.70,0.70,0.70}
\definecolor{grey71}{rgb}{0.71,0.71,0.71}
\definecolor{grey72}{rgb}{0.72,0.72,0.72}
\definecolor{grey73}{rgb}{0.73,0.73,0.73}
\definecolor{grey74}{rgb}{0.74,0.74,0.74}
\definecolor{grey75}{rgb}{0.75,0.75,0.75}
\definecolor{grey76}{rgb}{0.76,0.76,0.76}
\definecolor{grey77}{rgb}{0.77,0.77,0.77}
\definecolor{grey78}{rgb}{0.78,0.78,0.78}
\definecolor{grey79}{rgb}{0.79,0.79,0.79}
\definecolor{grey7}{rgb}{0.07,0.07,0.07}
\definecolor{grey80}{rgb}{0.80,0.80,0.80}
\definecolor{grey81}{rgb}{0.81,0.81,0.81}
\definecolor{grey82}{rgb}{0.82,0.82,0.82}
\definecolor{grey83}{rgb}{0.83,0.83,0.83}
\definecolor{grey84}{rgb}{0.84,0.84,0.84}
\definecolor{grey85}{rgb}{0.85,0.85,0.85}
\definecolor{grey86}{rgb}{0.86,0.86,0.86}
\definecolor{grey87}{rgb}{0.87,0.87,0.87}
\definecolor{grey88}{rgb}{0.88,0.88,0.88}
\definecolor{grey89}{rgb}{0.89,0.89,0.89}
\definecolor{grey8}{rgb}{0.08,0.08,0.08}
\definecolor{grey90}{rgb}{0.90,0.90,0.90}
\definecolor{grey91}{rgb}{0.91,0.91,0.91}
\definecolor{grey92}{rgb}{0.92,0.92,0.92}
\definecolor{grey93}{rgb}{0.93,0.93,0.93}
\definecolor{grey94}{rgb}{0.94,0.94,0.94}
\definecolor{grey95}{rgb}{0.95,0.95,0.95}
\definecolor{grey96}{rgb}{0.96,0.96,0.96}
\definecolor{grey97}{rgb}{0.97,0.97,0.97}
\definecolor{grey98}{rgb}{0.98,0.98,0.98}
\definecolor{grey99}{rgb}{0.99,0.99,0.99}
\definecolor{grey9}{rgb}{0.09,0.09,0.09}
\definecolor{grey}{rgb}{0.75,0.75,0.75}
\definecolor{honeydew1}{rgb}{0.94,1.00,0.94}
\definecolor{honeydew2}{rgb}{0.88,0.93,0.88}
\definecolor{honeydew3}{rgb}{0.76,0.80,0.76}
\definecolor{honeydew4}{rgb}{0.51,0.55,0.51}
\definecolor{honeydew}{rgb}{0.94,1.00,0.94}
\definecolor{hotpink}{rgb}{1.00,0.41,0.71}
\definecolor{indianred}{rgb}{0.80,0.36,0.36}
\definecolor{ivory1}{rgb}{1.00,1.00,0.94}
\definecolor{ivory2}{rgb}{0.93,0.93,0.88}
\definecolor{ivory3}{rgb}{0.80,0.80,0.76}
\definecolor{ivory4}{rgb}{0.55,0.55,0.51}
\definecolor{ivory}{rgb}{1.00,1.00,0.94}
\definecolor{khaki1}{rgb}{1.00,0.96,0.56}
\definecolor{khaki2}{rgb}{0.93,0.90,0.52}
\definecolor{khaki3}{rgb}{0.80,0.78,0.45}
\definecolor{khaki4}{rgb}{0.55,0.53,0.31}
\definecolor{khaki}{rgb}{0.94,0.90,0.55}
\definecolor{lavenderblush}{rgb}{1.00,0.94,0.96}
\definecolor{lavender}{rgb}{0.90,0.90,0.98}
\definecolor{lawngreen}{rgb}{0.49,0.99,0.00}
\definecolor{lemonchiffon}{rgb}{1.00,0.98,0.80}
\definecolor{lightblue}{rgb}{0.68,0.85,0.90}
\definecolor{lightcoral}{rgb}{0.94,0.50,0.50}
\definecolor{lightcyan}{rgb}{0.88,1.00,1.00}
\definecolor{lightgoldenrod}{rgb}{0.93,0.87,0.51}
\definecolor{lightgoldenrod}{rgb}{0.98,0.98,0.82}
\definecolor{lightgray}{rgb}{0.83,0.83,0.83}
\definecolor{lightgreen}{rgb}{0.56,0.93,0.56}
\definecolor{lightgrey}{rgb}{0.83,0.83,0.83}
\definecolor{lightpink}{rgb}{1.00,0.71,0.76}
\definecolor{lightsalmon}{rgb}{1.00,0.63,0.48}
\definecolor{lightsea}{rgb}{0.13,0.70,0.67}
\definecolor{lightsky}{rgb}{0.53,0.81,0.98}
\definecolor{lightslate}{rgb}{0.47,0.53,0.60}
\definecolor{lightslate}{rgb}{0.47,0.53,0.60}
\definecolor{lightslate}{rgb}{0.52,0.44,1.00}
\definecolor{lightsteel}{rgb}{0.69,0.77,0.87}
\definecolor{lightyellow}{rgb}{1.00,1.00,0.88}
\definecolor{limegreen}{rgb}{0.20,0.80,0.20}
\definecolor{linen}{rgb}{0.98,0.94,0.90}
\definecolor{magenta1}{rgb}{1.00,0.00,1.00}
\definecolor{magenta2}{rgb}{0.93,0.00,0.93}
\definecolor{magenta3}{rgb}{0.80,0.00,0.80}
\definecolor{magenta4}{rgb}{0.55,0.00,0.55}
\definecolor{magenta}{rgb}{1.00,0.00,1.00}
\definecolor{maroon1}{rgb}{1.00,0.20,0.70}
\definecolor{maroon2}{rgb}{0.93,0.19,0.65}
\definecolor{maroon3}{rgb}{0.80,0.16,0.56}
\definecolor{maroon4}{rgb}{0.55,0.11,0.38}
\definecolor{maroon}{rgb}{0.69,0.19,0.38}
\definecolor{mediumaquamarine}{rgb}{0.40,0.80,0.67}
\definecolor{mediumblue}{rgb}{0.00,0.00,0.80}
\definecolor{mediumorchid}{rgb}{0.73,0.33,0.83}
\definecolor{mediumpurple}{rgb}{0.58,0.44,0.86}
\definecolor{mediumsea}{rgb}{0.24,0.70,0.44}
\definecolor{mediumslate}{rgb}{0.48,0.41,0.93}
\definecolor{mediumspring}{rgb}{0.00,0.98,0.60}
\definecolor{mediumturquoise}{rgb}{0.28,0.82,0.80}
\definecolor{mediumviolet}{rgb}{0.78,0.08,0.52}
\definecolor{midnightblue}{rgb}{0.10,0.10,0.44}
\definecolor{mintcream}{rgb}{0.96,1.00,0.98}
\definecolor{mistyrose}{rgb}{1.00,0.89,0.88}
\definecolor{moccasin}{rgb}{1.00,0.89,0.71}
\definecolor{navajowhite}{rgb}{1.00,0.87,0.68}
\definecolor{navyblue}{rgb}{0.00,0.00,0.50}
\definecolor{navy}{rgb}{0.00,0.00,0.50}
\definecolor{oldlace}{rgb}{0.99,0.96,0.90}
\definecolor{olivedrab}{rgb}{0.42,0.56,0.14}
\definecolor{orange1}{rgb}{1.00,0.65,0.00}
\definecolor{orange2}{rgb}{0.93,0.60,0.00}
\definecolor{orange3}{rgb}{0.80,0.52,0.00}
\definecolor{orange4}{rgb}{0.55,0.35,0.00}
\definecolor{orangered}{rgb}{1.00,0.27,0.00}
\definecolor{orange}{rgb}{1.00,0.65,0.00}
\definecolor{orchid1}{rgb}{1.00,0.51,0.98}
\definecolor{orchid2}{rgb}{0.93,0.48,0.91}
\definecolor{orchid3}{rgb}{0.80,0.41,0.79}
\definecolor{orchid4}{rgb}{0.55,0.28,0.54}
\definecolor{orchid}{rgb}{0.85,0.44,0.84}
\definecolor{palegoldenrod}{rgb}{0.93,0.91,0.67}
\definecolor{palegreen}{rgb}{0.60,0.98,0.60}
\definecolor{paleturquoise}{rgb}{0.69,0.93,0.93}
\definecolor{paleviolet}{rgb}{0.86,0.44,0.58}
\definecolor{papayawhip}{rgb}{1.00,0.94,0.84}
\definecolor{peachpuff}{rgb}{1.00,0.85,0.73}
\definecolor{peru}{rgb}{0.80,0.52,0.25}
\definecolor{pink1}{rgb}{1.00,0.71,0.77}
\definecolor{pink2}{rgb}{0.93,0.66,0.72}
\definecolor{pink3}{rgb}{0.80,0.57,0.62}
\definecolor{pink4}{rgb}{0.55,0.39,0.42}
\definecolor{pink}{rgb}{1.00,0.75,0.80}
\definecolor{plum1}{rgb}{1.00,0.73,1.00}
\definecolor{plum2}{rgb}{0.93,0.68,0.93}
\definecolor{plum3}{rgb}{0.80,0.59,0.80}
\definecolor{plum4}{rgb}{0.55,0.40,0.55}
\definecolor{plum}{rgb}{0.87,0.63,0.87}
\definecolor{powderblue}{rgb}{0.69,0.88,0.90}
\definecolor{purple1}{rgb}{0.61,0.19,1.00}
\definecolor{purple2}{rgb}{0.57,0.17,0.93}
\definecolor{purple3}{rgb}{0.49,0.15,0.80}
\definecolor{purple4}{rgb}{0.33,0.10,0.55}
\definecolor{purple}{rgb}{0.63,0.13,0.94}
\definecolor{red1}{rgb}{1.00,0.00,0.00}
\definecolor{red2}{rgb}{0.93,0.00,0.00}
\definecolor{red3}{rgb}{0.80,0.00,0.00}
\definecolor{red4}{rgb}{0.55,0.00,0.00}
\definecolor{red}{rgb}{1.00,0.00,0.00}
\definecolor{rosybrown}{rgb}{0.74,0.56,0.56}
\definecolor{royalblue}{rgb}{0.25,0.41,0.88}
\definecolor{saddlebrown}{rgb}{0.55,0.27,0.07}
\definecolor{salmon1}{rgb}{1.00,0.55,0.41}
\definecolor{salmon2}{rgb}{0.93,0.51,0.38}
\definecolor{salmon3}{rgb}{0.80,0.44,0.33}
\definecolor{salmon4}{rgb}{0.55,0.30,0.22}
\definecolor{salmon}{rgb}{0.98,0.50,0.45}
\definecolor{sandybrown}{rgb}{0.96,0.64,0.38}
\definecolor{seagreen}{rgb}{0.18,0.55,0.34}
\definecolor{seashell1}{rgb}{1.00,0.96,0.93}
\definecolor{seashell2}{rgb}{0.93,0.90,0.87}
\definecolor{seashell3}{rgb}{0.80,0.77,0.75}
\definecolor{seashell4}{rgb}{0.55,0.53,0.51}
\definecolor{seashell}{rgb}{1.00,0.96,0.93}
\definecolor{sienna1}{rgb}{1.00,0.51,0.28}
\definecolor{sienna2}{rgb}{0.93,0.47,0.26}
\definecolor{sienna3}{rgb}{0.80,0.41,0.22}
\definecolor{sienna4}{rgb}{0.55,0.28,0.15}
\definecolor{sienna}{rgb}{0.63,0.32,0.18}
\definecolor{skyblue}{rgb}{0.53,0.81,0.92}
\definecolor{slateblue}{rgb}{0.42,0.35,0.80}
\definecolor{slategray}{rgb}{0.44,0.50,0.56}
\definecolor{slategrey}{rgb}{0.44,0.50,0.56}
\definecolor{snow1}{rgb}{1.00,0.98,0.98}
\definecolor{snow2}{rgb}{0.93,0.91,0.91}
\definecolor{snow3}{rgb}{0.80,0.79,0.79}
\definecolor{snow4}{rgb}{0.55,0.54,0.54}
\definecolor{snow}{rgb}{1.00,0.98,0.98}
\definecolor{springgreen}{rgb}{0.00,1.00,0.50}
\definecolor{steelblue}{rgb}{0.27,0.51,0.71}
\definecolor{tan1}{rgb}{1.00,0.65,0.31}
\definecolor{tan2}{rgb}{0.93,0.60,0.29}
\definecolor{tan3}{rgb}{0.80,0.52,0.25}
\definecolor{tan4}{rgb}{0.55,0.35,0.17}
\definecolor{tan}{rgb}{0.82,0.71,0.55}
\definecolor{thistle1}{rgb}{1.00,0.88,1.00}
\definecolor{thistle2}{rgb}{0.93,0.82,0.93}
\definecolor{thistle3}{rgb}{0.80,0.71,0.80}
\definecolor{thistle4}{rgb}{0.55,0.48,0.55}
\definecolor{thistle}{rgb}{0.85,0.75,0.85}
\definecolor{tomato1}{rgb}{1.00,0.39,0.28}
\definecolor{tomato2}{rgb}{0.93,0.36,0.26}
\definecolor{tomato3}{rgb}{0.80,0.31,0.22}
\definecolor{tomato4}{rgb}{0.55,0.21,0.15}
\definecolor{tomato}{rgb}{1.00,0.39,0.28}
\definecolor{turquoise1}{rgb}{0.00,0.96,1.00}
\definecolor{turquoise2}{rgb}{0.00,0.90,0.93}
\definecolor{turquoise3}{rgb}{0.00,0.77,0.80}
\definecolor{turquoise4}{rgb}{0.00,0.53,0.55}
\definecolor{turquoise}{rgb}{0.25,0.88,0.82}
\definecolor{violetred}{rgb}{0.82,0.13,0.56}
\definecolor{violet}{rgb}{0.93,0.51,0.93}
\definecolor{wheat1}{rgb}{1.00,0.91,0.73}
\definecolor{wheat2}{rgb}{0.93,0.85,0.68}
\definecolor{wheat3}{rgb}{0.80,0.73,0.59}
\definecolor{wheat4}{rgb}{0.55,0.49,0.40}
\definecolor{wheat}{rgb}{0.96,0.87,0.70}
\definecolor{whitesmoke}{rgb}{0.96,0.96,0.96}
\definecolor{white}{rgb}{1.00,1.00,1.00}
\definecolor{yellow1}{rgb}{1.00,1.00,0.00}
\definecolor{yellow2}{rgb}{0.93,0.93,0.00}
\definecolor{yellow3}{rgb}{0.80,0.80,0.00}
\definecolor{yellow4}{rgb}{0.55,0.55,0.00}
\definecolor{yellowgreen}{rgb}{0.60,0.80,0.20}
\definecolor{yellow}{rgb}{1.00,1.00,0.00}
\usepackage{amsfonts}
\usepackage{bm}
\usepackage{longtable}
\usepackage{pifont}
\usepackage{booktabs} 
 
\newcommand{\whofont}{
  \fontfamily{pcr}
  \bfseries 
  \color{royalblue}
}

\newcommand{\gskfont}{
  \fontfamily{pcr}
  \bfseries 
  \color{mediumorchid}
}
\newcommand{\jcafont}{
  \color{applered}
}
\newcommand{\mcfont}{
  \fontfamily{pcr}
  \bfseries 
  \color{azure}
}

\DeclareTextFontCommand{\who}{\whofont}
\DeclareTextFontCommand{\gsk}{\gskfont}
\DeclareTextFontCommand{\jca}{\jcafont}
\DeclareTextFontCommand{\mc}{\mcfont}

\shorttitle{Mg~\textsc{ii} Non-Equilibrium}
\shortauthors{Kerr, Carlsson, Allred}

\begin{document}

%%%%%%%%%%%%%%%%%%%%%%%%%%%%%%%%%%%%%%%%%%%%%%%%%%%%%
%%%%%%%%%%%%%%%%%%%%     TITLE & ABSTRACT     %%%%%%%%%%%%%%%%%%%%
%%%%%%%%%%%%%%%%%%%%%%%%%%%%%%%%%%%%%%%%%%%%%%%%%%%%%

	\title{Modelling Mg~\textsc{ii} During Solar Flares, II: Non-equilibrium Effects}
	
	\author{Graham~S.~Kerr}
	\email{graham.s.kerr@nasa.gov}
	\altaffiliation{NPP Fellow, administered by USRA}
	\affil{NASA Goddard Space Flight Center, Heliophysics Sciences Division, Code 671, 8800 Greenbelt Rd., Greenbelt, MD 20771, USA}
 	
	 \author{Mats Carlsson}
	 \affil{Rosseland Centre for Solar Physics, University of Oslo, P.O. Box 1029, Blindern, N-0315 Oslo, Norway}
	 \affil{Institute of Theoretical Astrophysics, University of Oslo, P.O. Box 1029, Blindern, N-0315 Oslo, Norway}
	  	   
	 \author{Joel~C.~Allred}
	 \affil{NASA Goddard Space Flight Center, Heliophysics Sciences Division, Code 671, 8800 Greenbelt Rd., Greenbelt, MD 20771, USA}
	  			 
	\date{Received / Accepted}
	
	\keywords{Sun: chromosphere  - Sun: flares - Sun: UV radiation - radiative transfer - methods: numerical}
	
	\begin{abstract}	
	To extract the information that the Mg~\textsc{ii} NUV spectra (observed by the Interface Region Imaging Spectrograph; IRIS), carries about the chromosphere during solar flares, and to validate models of energy transport via model-data comparison, forward modelling is required. The assumption of statistical equilibrium is typically used to obtain the atomic level populations from snapshots of flare atmospheres, due to computational necessity. However it is possible that relying on statistical equilibrium could lead to spurious results. We compare solving the atomic level populations via statistical equilibrium versus a non-equilibrium time-dependent approach. This was achieved using flare simulations from \texttt{RADYN} alongside the minority species version, \texttt{MS\_RADYN}, from which the time-dependent Mg~\textsc{ii} atomic level populations and radiation transfer were computed in complete frequency redistribution. The impacts on the emergent profiles, lightcurves, line ratios, and formation heights are discussed. In summary we note that non-equilibrium effects during flares are typically important only in the initial stages and for a short period following the cessation of the energy injection. An analysis of the timescales of ionisation equilibrium reveals that for most of the duration of the flare, when the temperatures and densities are sufficiently enhanced, the relaxation timescales are short ($\tau_{\mathrm{relax}}<0.1$ s), so that the equilibrium solution is an adequate approximation. These effects vary with the size of the flare, however. In weaker flares effects can be more pronounced. We recommend that non-equilibrium effects be considered when possible, but that statistical equilibrium is sufficient at most stages of the flare.		   
	\end{abstract}

%%%%%%%%%%%%%%%%%%%%%%%%%%%%%%%%%%%%%%%%%%%%%%%%%%%%%
%%%%%%%%%%%%%%%%%%%%     INTRODUCTION      %%%%%%%%%%%%%%%%%%%%%
%%%%%%%%%%%%%%%%%%%%%%%%%%%%%%%%%%%%%%%%%%%%%%%%%%%%%

\section{Introduction}\label{sec:intro}
Solar flares, and other transient energy release events, can dramatically disturb the solar atmosphere, driving it out of equilibrium. The importance of non-equilibrium (NEQ) ionisation and recombination in determining the atomic level populations (and therefore the radiative response to the disturbance) will likely vary from species to species, and the charge states of those species. However, when modelling the radiative response of the atmosphere to flares it is often the case that the statistical equilibrium (SE) solution is sought, given the computational demands of performing time dependent simulations that include physical processes, such as partial frequency redistribution, that are required for certain transitions \citep[recent examples include][]{2017ApJ...842...82R,2019ApJ...879...19Z,kerr_2019}. This is partly mitigated by using the non-equilibrium electron density from dynamic simulations. We address the importance of non-equilibrium ionisation on the modelling of Mg~\textsc{ii} h \& k lines during solar flares, a routinely observed pair of strong spectral lines from the Sun.  

In the standard flare model, magnetic reconnection occurs in the corona, releasing vast amounts of magnetic energy, up to $10^{32}$~ergs, \citep{2011SSRv..159...19F} . This results in \textsl{in situ} heating and particle acceleration. Electrons (and likely ions) are accelerated at relativistic speeds along flare loops, loosing energy via Coulomb interactions as they travel into denser regions \citep[][]{1971SoPh...18..489B,2011SSRv..159..107H,2019ApJ...880..136J}. A broadband enhancement to the solar radiative output results from both thermal and non-thermal processes following plasma heating and ionisation. Chromospheric ablations (`evaporations'), with bulk upflows reaching several hundreds of km s$^{-1}$, and condensations, with bulk downflows of dense material reaching a few tens of km s$^{-1}$, are also driven during flares, revealed by Doppler shifts of spectral lines \citep[e.g.][]{1985ApJ...289..414F,1989ApJ...346.1019F,2009ApJ...699..968M,2015ApJ...807L..22G}.

The transition region and chromosphere are both the sites of energy deposition and the origin of the bulk of the radiative output during flares \citep{2011SSRv..159...19F,2014ApJ...793...70M}. They are crucial locations for both diagnosing the flaring plasma, and as test grounds for models of energy transport during flares. For the former, forward modelling radiation from different flare atmospheres allows us to understand how to extract physical properties of the flaring plasma from observations. For the latter, model-data comparisons allows us to understand how well simulated flare atmospheres driven by a particular energy transport mechanism compare with the actual flaring chromosphere, and therefore how consistent our models of energy transport are with the reality.

Since the launch of the Interface Region Imaging Spectrograph \citep[][IRIS]{2014SoPh..289.2733D}, observations of the Mg~\textsc{ii} h \& k resonance and subordinate lines have become commonly used in observational studies \citep[e.g.][]{2015A&A...582A..50K,2015SoPh..290.3525L,2015ApJ...807L..22G,2018PASJ...70..100T,2018ApJ...856...34T,2018ApJ...861...62P}, and are attractive as a means to diagnose or critically attack flare (and other) models \citep[e.g][]{2016ApJ...827...38R,2016ApJ...827..101K,2019ApJ...879...19Z}. These lines have been observed in hundreds of flare events since the launch of IRIS, show significant variation from the quiet Sun during flare heating, and form over a range of chromospheric layers meaning they have the potential to diagnose a large part of the lower solar atmosphere. Particularly if they are combined with other IRIS or ground based observables. 

These lines are optically thick, and as such are complex to interpret, requiring forward modelling to understand their formation \citep{2013ApJ...772...89L}. Modelling of the Mg~\textsc{ii} NUV spectrum has been a key complementary activity to the analysis of observations, and it is essential that we are modelling these lines in an accurate manner. This is both to provide an accurate physical interpretation of these complex optically thick lines, but also to facilitate high fidelity model-data comparison by which models of flare energy transport are critically interrogated. 

In Paper 1 \citep{kerr_2019} we used \texttt{RADYN} \citep{1992ApJ...397L..59C,1997ApJ...481..500C,2015ApJ...809..104A} radiation hydrodynamic flare atmospheres with the radiation transport code \texttt{RH}, \citep{2001ApJ...557..389U}, a common approach to modelling these lines in flares. This was to investigate how various setups affect the Mg~\textsc{ii} solution such as partial frequency redistribution (PRD) vs complete frequency redistribution (CRD), inclusion of other species, and effects of coronal irradiation.  In particular we noted that PRD is still required, resulting in substantial differences to the CRD solution. However, in that work the level populations were obtained under the assumption of statistical equilibrium, since RH is a stationary code. Each atmospheric snapshot was treated in isolation, and the statistical equilibrium solution obtained. This was necessary in order to include the more advanced radiation transfer from the RH code (in particular PRD), that would be computationally very demanding to include in a time-dependent flare simulation. The history of the atmosphere and any time dependent effects were neglected. In this paper we will investigate the impact of non-equilibrium processes on the formation of Mg~\textsc{ii} during solar flares, with the aim of determining if the usual practice of omitting these effects is safe. 

The chromosphere is dynamic, especially during solar flares. \cite{1999AIPC..471...23C,2002ApJ...572..626C} demonstrated, from \texttt{RADYN} simulations of propagating acoustic waves, that the ionisation and recombination timescale ($\tau\sim10^3-10^5$~s) for hydrogen is long compared to the dynamical timescale, and that if statistical equilibrium is assumed then the ionisation fraction is underestimated by several orders of magnitude in certain locations of the atmosphere. The electron density would consequently be very different.  Similarly, the requirements for considering non-equilibrium effects on other species have been noted, such as Helium \citep{2014ApJ...784...30G,2016ApJ...817..125G}, O~\textsc{iv} \citep{2013ApJ...767...43O}, and Si~\textsc{iv} \citep{2016ApJ...817...46M}. \cite{2013ApJ...772...89L} investigated if non-equilibrium processes were important for Mg~\textsc{ii} in the quiet Sun, concluding that whenever the temperature was large enough to produce significant amounts of Mg~\textsc{iii}, the relaxation time was shortened, and that using statistical equilibrium was sufficient. Of course flare chromospheres are very dynamic and the conclusions of \cite{2013ApJ...772...89L} might not apply in flaring conditions. We investigate the flaring scenario in this work. 

The non-equilibrium atomic level population equation is: 

\begin{equation}\label{eq:nequil}
	\frac{\partial n_{i}}{\partial t} + \frac{\partial n_{i}v}{\partial z} - \left(\sum^{N^\prime}_{j\ne i} n_{j}P_{j,i} - n_{i}\sum^{N^{\prime}}_{i\ne j} P_{i,j} \right) = 0,
\end{equation}

\noindent where $v$ is the atmospheric velocity, $N^{\prime}$ is the total number of states, $t$ is time, $z$ is the height in the atmosphere, $P_{i,j}$ describes the total rates (collisional plus radiative) from $i$ to $j$, and $P_{j,i}$ is the total rate from $j$ to $i$. The transition rates are functions of the local atmospheric conditions (including energy input) which vary with time in dynamical simulations. If the local thermodynamic state of the atmosphere or the radiation field vary faster than the timescale for ionisation and recombination then there is not enough time for the atmosphere to reach equilibrium - the `history' of the atmosphere becomes important. 

If the ionisation and recombination timescales are sufficiently fast then the populations can be approximated by statistical equilibrium (setting $\partial n_{i}/\partial t$ and $\partial n_{i}v/\partial z$ to zero in Equation~\ref{eq:nequil}):

\begin{equation}\label{eq:statequil}
	 \sum^{N^\prime}_{j\ne i} n_{j}P_{j,i}  =  n_{i}\sum^{N^{\prime}}_{i\ne j} P_{i,j} ,
\end{equation}

\noindent In situations where the history of the atmosphere is important but statistical equilibrium is used, there may consequently be errors in the population densities of atomic states, and in the synthetic spectra.

%%%%%%%%%%%%%%%%%%%%%%%%%%%%%%%%%%%%%%%%%%%%%%%%%%%%%
%%%%%%%%%%%%%%%     NUMERICAL RESOURCES      %%%%%%%%%%%%%%%%%%%%%
%%%%%%%%%%%%%%%%%%%%%%%%%%%%%%%%%%%%%%%%%%%%%%%%%%%%%
\section{Flare Simulations}\label{sec:codes}
\begin{figure*}
	\centering 
	{\includegraphics[width = .95\textwidth, clip = true, trim = 0.cm 0.cm 0.cm 0.cm]{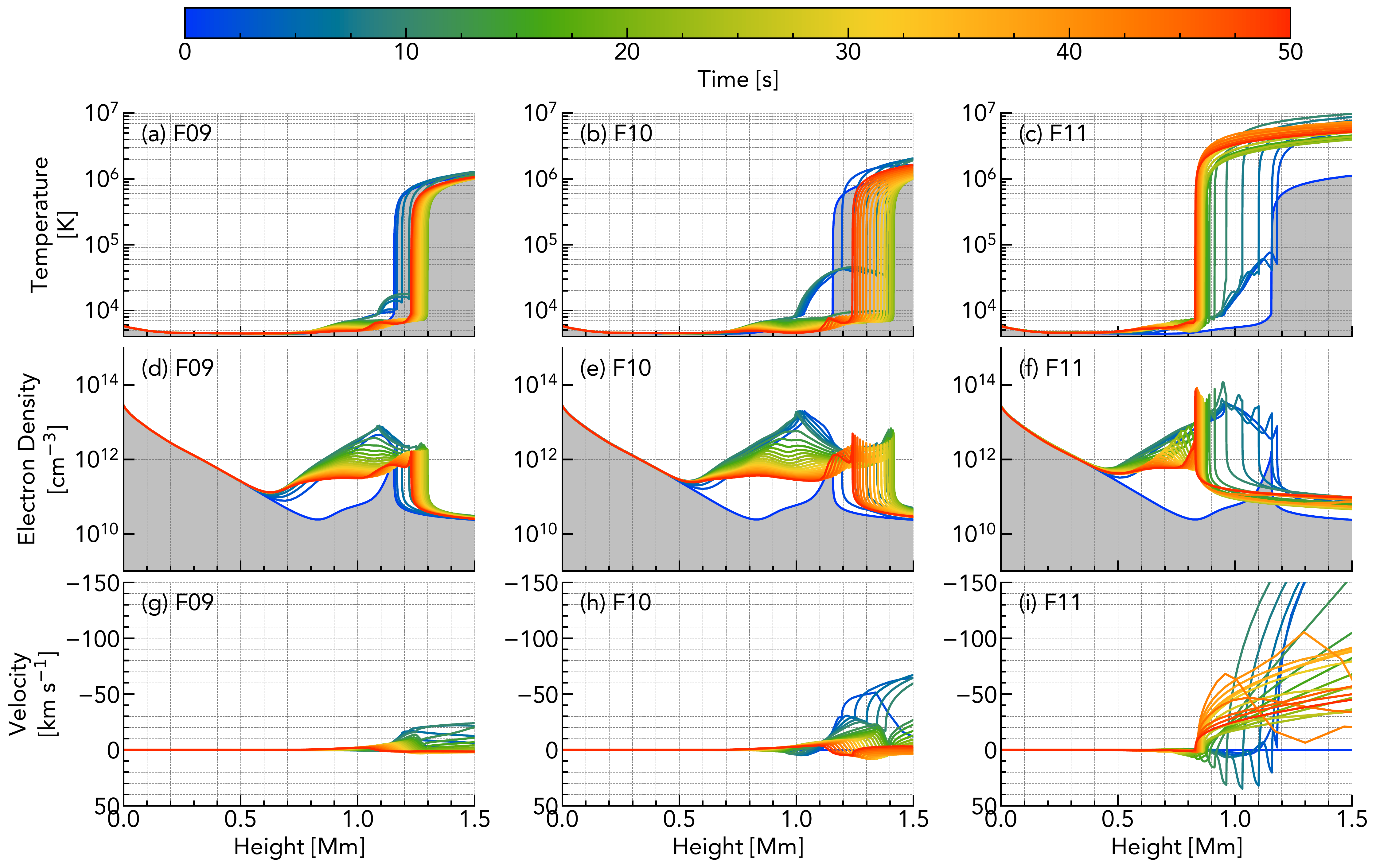}}	
	\caption{\textsl{Stratification of temperature (a,b,c), electron density (d,e,f), and macroscopic velocity (g,h,i; uplows are negative) in the three flare simulations. The first column shows the F9 simulation, second column shows F10 simulation, and third column shows the F11 simulation. Colour represents time. Recall that heating ceased at $t=10$~s}. Note that this is a reproduction of Figure~1 in \cite{kerr_2019}.}
	\label{fig:flare_atmos}
\end{figure*}
Three flare simulations were produced using the \texttt{RADYN} radiation hydrodynamic code \citep{1992ApJ...397L..59C,1997ApJ...481..500C,1999ApJ...521..906A,2005ApJ...630..573A,2015ApJ...809..104A,FP_inprep}, with the flares driven by non-thermal electron beams that were injected into a pre-flare atmosphere that spanned the sub-photosphere through to the corona. Both the code and details of these simulations are described in Paper 1 \citep{kerr_2019}. The electron beam fluxes were $F=[1\times10^{9}, 1\times10^{10}, 1\times10^{11}]$~erg~cm$^{-2}$~s$^{-1}$ (hereafter F9, F10, \& F11) which were modelled as power law distributions with spectra index $\delta = 5$ and a low-energy cutoff $E_{c} = 20$~keV. Energy was injected at a constant rate for $t=10$~s, and the simulations allowed to continue to evolve until $t=50$~s. Figure~\ref{fig:flare_atmos} shows the evolution of the atmospheres as a function of time in each simulations. The grey shaded portion is the pre-flare atmosphere. 

\texttt{RADYN} simulated the time-dependent response of the atmospheres to flare energy injection so that the atomic level populations, electron density, and temperature stratification are all NEQ, but with radiation computed using the simplifying assumption of CRD (necessary to make the dynamic flare problem computationally tractable). Species important for energy balance are considered in the main simulation (H, He, and Ca~\textsc{ii}), with other species included in an optically thin radiation loss function. While Mg~\textsc{ii} h \& k are very strong chromospheric lines, PRD is required to accurately model the radiative losses. Omitting the Mg~\textsc{ii} h \& k lines is likely safe to do from an energetic balance standpoint, given that the Ca~\textsc{ii}  H \& K lines \textsl{are} included, but in CRD. So, losses from Ca~\textsc{ii} H \& K are overestimated, but this is mitigated by ignoring Mg~\textsc{ii} h \& k. % It was also shown in \cite{kerr_2019} that a sufficiently large atomic model is required for Mg~\textsc{ii} modelling, which would adversely affect the \texttt{RADYN} runtime.

In order to obtain the Mg~\textsc{ii} NEQ atomic level populations and emergent profiles we must therefore turn to \texttt{MS\_RADYN}, the minority species version of \texttt{RADYN}. This code uses each internal timestep of an existing \texttt{RADYN} RHD solution to solve the NEQ NLTE radiation transport problem for a desired minority species, such as Mg~\textsc{ii}. It includes the time dependent and advection terms when solving the atomic level populations. Since the variables required for this simulation are stored for each internal timestep (and not the output cadence of the main set of variables, which can be many times larger) \texttt{MS\_RADYN} can capture changes to the atomic level populations on very short timescales that result from the changing atmospheric state.  \texttt{MS\_RADYN} was used in this fashion to study C~\textsc{ii} emission \citep{2003ApJ...597.1158J} and more recently to investigate radiative transfer effects on Si~\textsc{iv} during flares \citep{2019ApJ...871...23K}.

Our three flare simulations were used as input in \texttt{MS\_RADYN} to simulate the atomic level populations and synthetic spectra of a ten-level-with-continuum model of Mg~\textsc{ii}, the same as used in \cite{2013ApJ...772...89L} and that we used in \cite{kerr_2019}. This model atom included the h \& k transitions, and the subordinate line transitions. To test the impact of NEQ effects we then repeated these simulations, but switched off the time-dependent and advection terms when computing the atomic level populations so that \texttt{MS\_RADYN} then used statistical equilibrium (Equation~\ref{eq:statequil}). In this latter series of simulations, the Mg~\textsc{ii} problem was solved using the non-equilibrium hydrogen and electron densities (similar to the more typical post-processing of RHD/HD snapshots through \texttt{RH}). 

Using \texttt{MS\_RADYN} in this fashion meant that the only differences were the terms included in the atomic level population equation (Equation~\ref{eq:nequil} vs \ref{eq:statequil}), with the same flare atmospheres and background opacities used throughout. Any differences in the level populations and emergent spectra are then due to non-equilibrium processes. 

%A necessary caveat here is that we are omitting potential blends and employ complete frequency redistribution. We do not anticipate that PRD would change any impact of non-equilibrium processes, and found that the atomic level populations were not strongly impacted by PRD in \cite{kerr_2019}. 

%%%%%%%%%%%%%%%%%%%%%%%%%%%%%%%%%%%%%%%%%%%%%%%%%%%%%
%%%%%%%%%%%%%%%%%%%%     LINE PROFILES      %%%%%%%%%%%%%%%%%%%%%
%%%%%%%%%%%%%%%%%%%%%%%%%%%%%%%%%%%%%%%%%%%%%%%%%%%%%
\section{Line Profiles}\label{sec:emergent_intensity}

\begin{figure}
	\centering 
	{\includegraphics[width = 0.5\textwidth, clip = true, trim = 0.cm 0.cm 0.cm 0.cm]{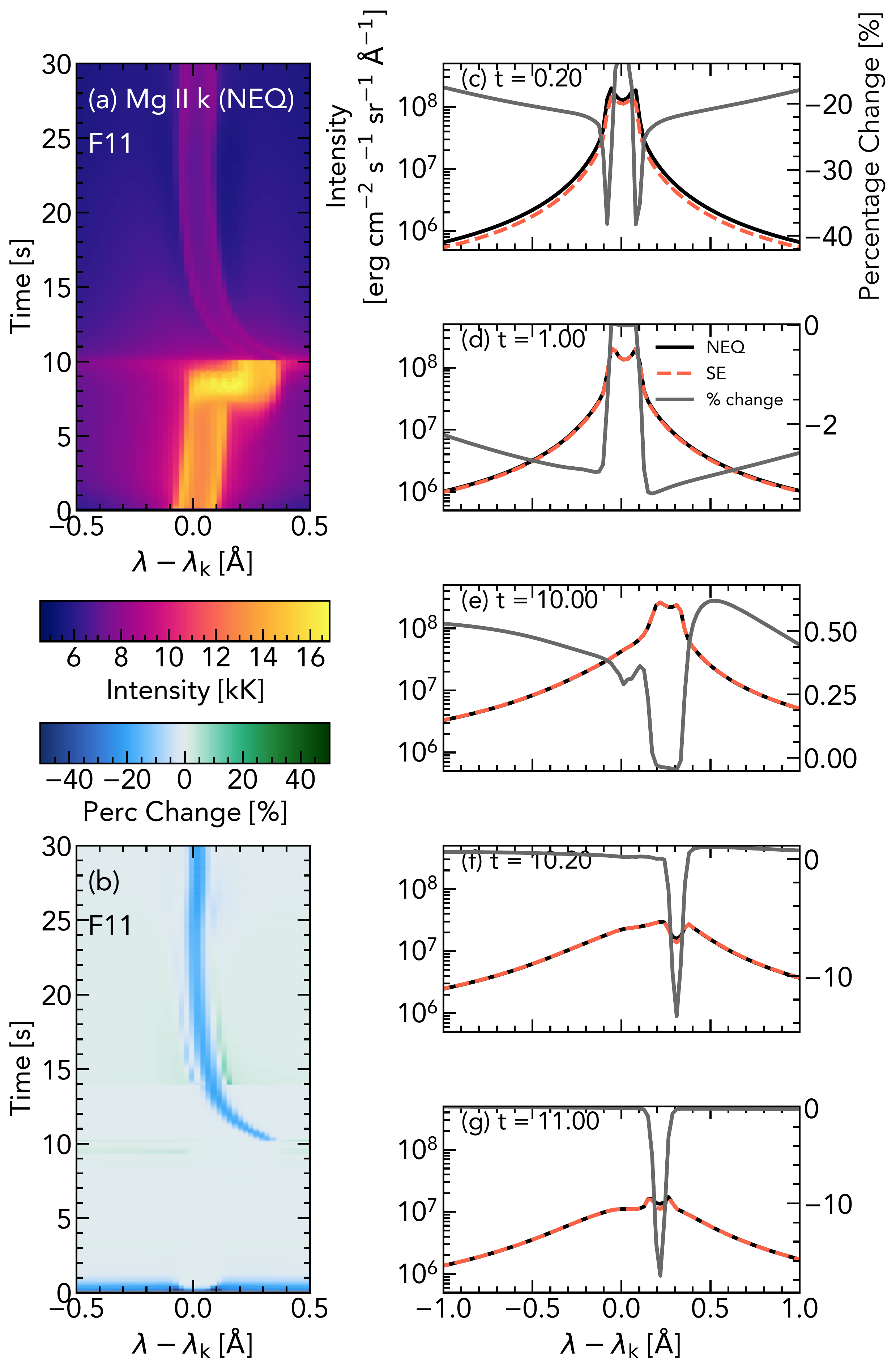}}	
	\caption{\textsl{Mg~\textsc{ii} k line profiles in the F11 simulation. In panels (a,b) wavelength is shown on the x-axis, and time is stacked on the y-axis, so that the temporal evolution is shown. Panel (a) shows the NEQ profiles, and panel (b) the percentage change between the NEQ and SE profiles (positive means that SE is more intense, negative that NEQ is more intense, and the change is saturated on the scale indicated). Panels (c-g) show a comparison of the NEQ (black lines) to SE (red dashed lines) line profiles at selected times, where the grey lines are the percentage change.}}
	\label{fig:f11_kline_stack}
\end{figure}
\begin{figure}
	\centering 
	{\includegraphics[width = 0.5\textwidth, clip = true, trim = 0.cm 0.cm 0.cm 0.cm]{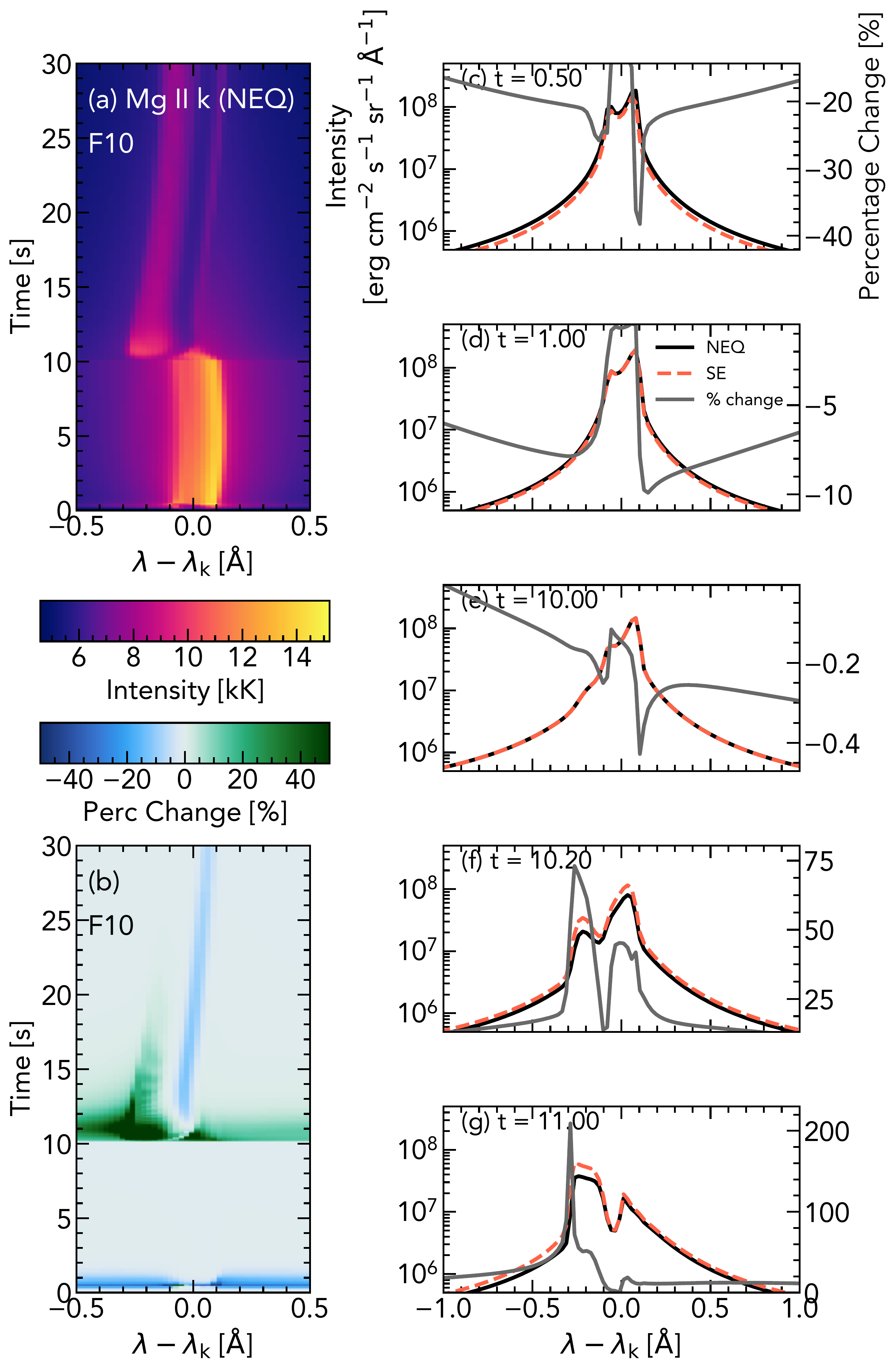}}	
	\caption{\textsl{Same as Figure~\ref{fig:f11_kline_stack}, but for the F10 simulation.}}
	\label{fig:f10_kline_stack}
\end{figure}
\begin{figure}
	\centering 
	{\includegraphics[width = 0.5\textwidth, clip = true, trim = 0.cm 0.cm 0.cm 0.cm]{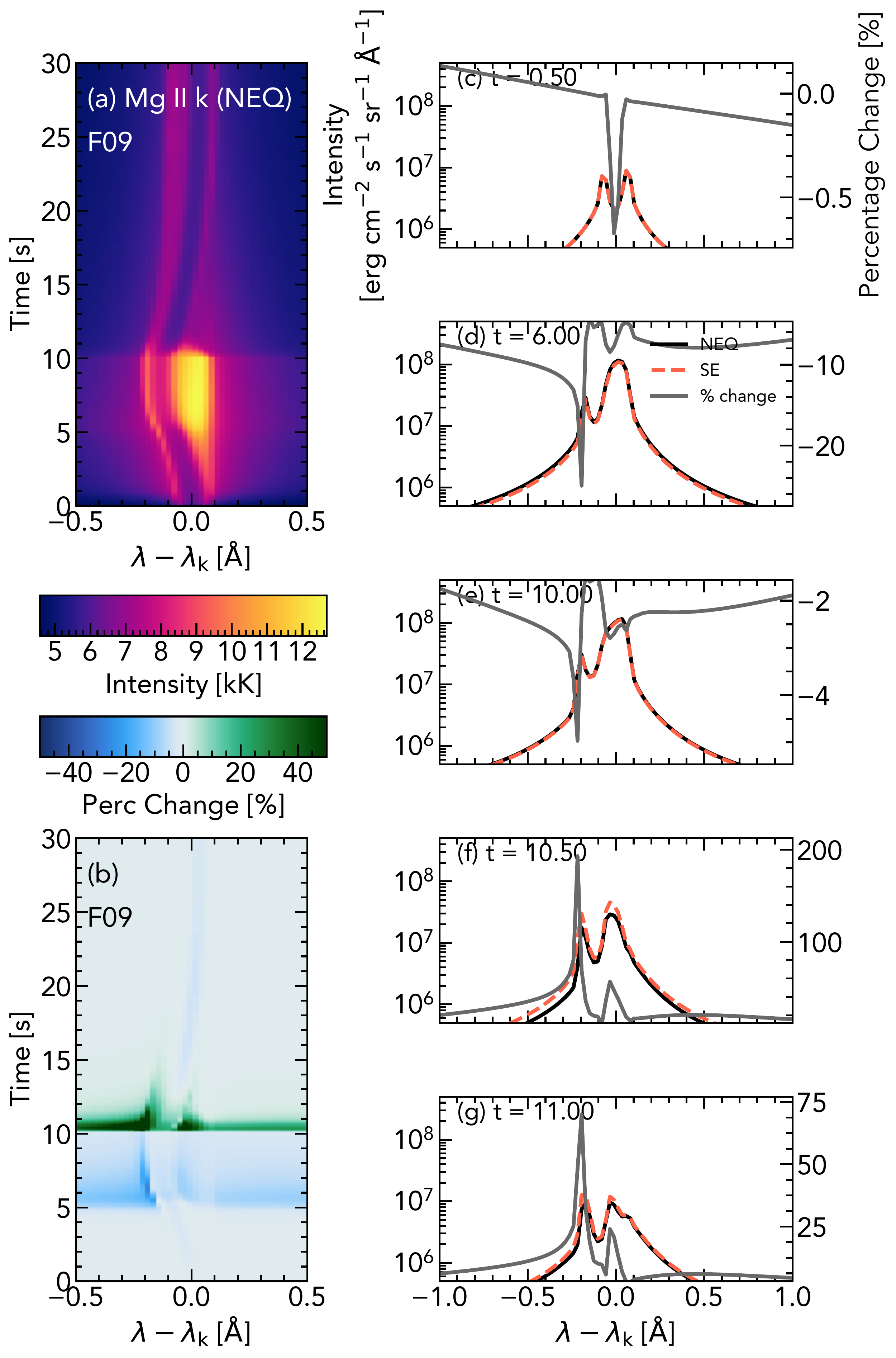}}	
	\caption{\textsl{\textsl{Same as Figure~\ref{fig:f11_kline_stack}, but for the F9 simulation}}}
	\label{fig:f09_kline_stack}
\end{figure}
\begin{figure}
	\centering 
	{\includegraphics[width = 0.5\textwidth, clip = true, trim = 0.cm 0.cm 0.cm 0.cm]{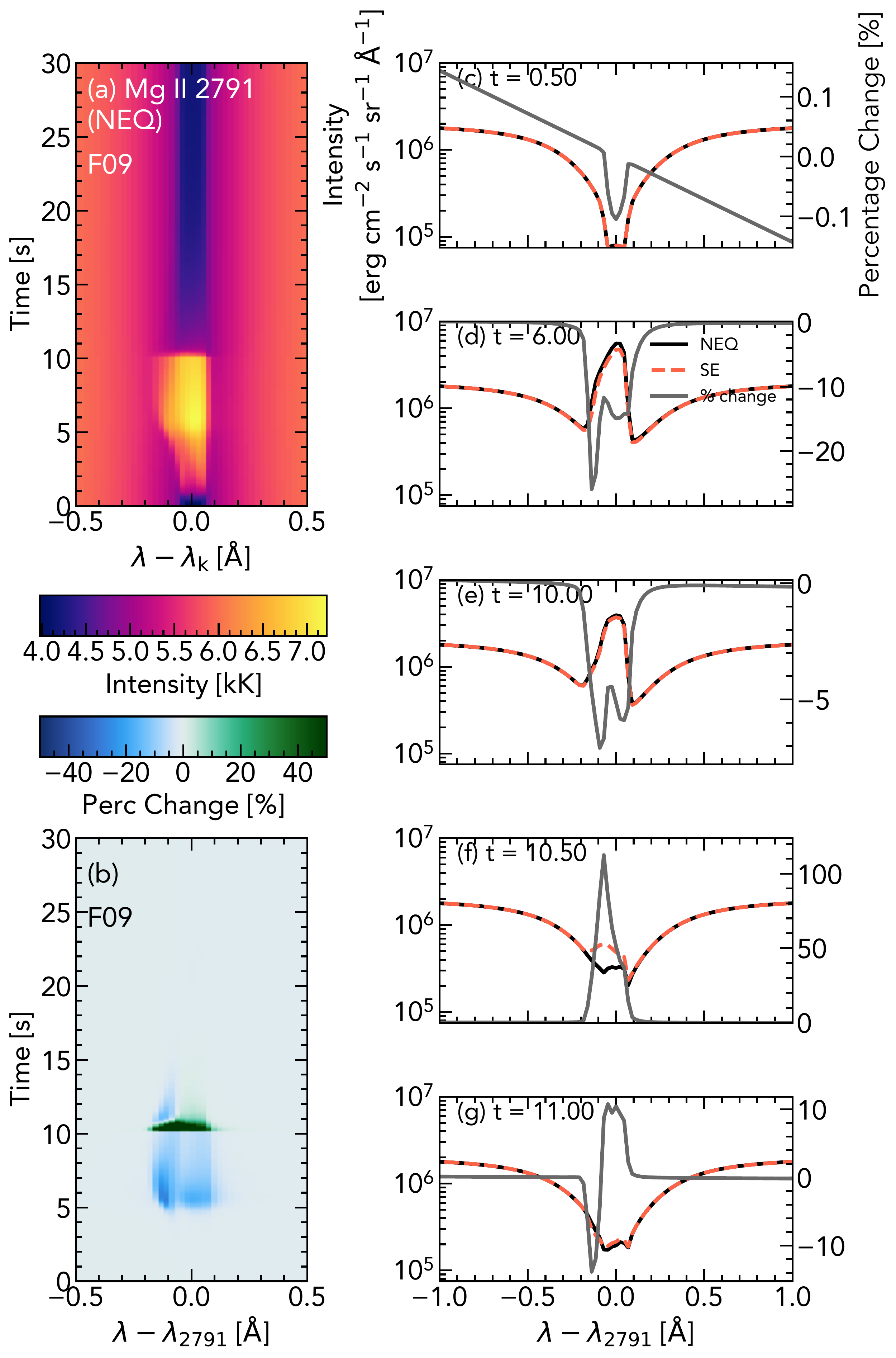}}	
	\caption{\textsl{\textsl{Same as Figure~\ref{fig:f11_kline_stack}, but for the F9 simulation, and here we show the Mg~\textsc{ii} 2791\AA\ subordinate line.}}}
	\label{fig:f09_2791_stack}
\end{figure}

How the emergent profiles differ between each simulation is the most important consideration indicating if there is a negligible, tolerable, or significant difference between NEQ and SE when modelling Mg~\textsc{ii}, given our goal of comparing to observations. Here we discuss these differences in each scenario, but we do not address in detail the line features themselves and how they vary with flare strength or how well they compare to observations, which will be the focus of other investigations. 
 
To quantify the difference between the two results we compute the percentage change between the intensities across the resonance and subrodinate line profiles, $\frac{I_{\mathrm{SE}} - I_{\mathrm{NEQ}}}{I_{\mathrm{NEQ}}}\times100$, where $I_{\mathrm{SE}}$ is the intensity computed using SE, and $I_{\mathrm{NEQ}}$ is the intensity computed using NEQ. A positive percentage means that $I_{\mathrm{SE}}$ is more intense.

For the bulk of the duration of the flares the differences between the results is only minor or negligible, though there are exceptions. Figures~\ref{fig:f11_kline_stack},~\ref{fig:f10_kline_stack}, and \ref{fig:f09_kline_stack} show a comparison between the NEQ and SE Mg~\textsc{ii} k line profiles, and the percentage change between the two solutions, for the F11, F10, and F9 simulations respectively. The Mg~\textsc{ii} 2791~\AA\ line is shown for the F9 simulation in Figure~\ref{fig:f09_2791_stack}. In each figure panel (a) shows the line profile as a function of time (stacked on the y-axis), and panel (b) shows the percentage change. Panels (c-g) show profiles at selected times of interest. 

It is immediately clear from these figures that while assuming SE results in very little change during the main heating phase, it does result in under- or over-estimations of the line intensity across the whole k line during the initial energy injection and following cessation of the beam. The subordinate line differences are more limited to the line core and near wings. Line shapes and features are preserved, though in the decay phase the subordinate lines return from emission to absorption more rapidly in the NEQ simulations. 

Within the first second of the F11 and F10 simulations, there is an intensity change $-[5-50]\%$, with the strongest changes around the emission peaks. These differences rapidly subside in both F11 and F10, until the cessation of energy input after which a very narrow region around the line core shows approximately a $10-20\%$ change, again with NEQ more intense. In the F10 simulation there are several seconds following the cessation of energy input where the NEQ solution is significantly less intense, up to approximately $230\%$ in the blue emission peak. The F9 simulation evolves more gradually, and shows close agreement between the NEQ and SE solutions in the initial stages. Unlike the other two simulations, however, differences appear during the heating phase, with a larger NEQ result, with a difference on the order $5-20\%$. In the cooling phase the F9 behaves similar to the F10 simulation, with a stronger SE intensity. The h \& k lines are not affected by the same magnitude meaning the k:h line ratio will vary between the NEQ and SE solutions, as discussed in Section~\ref{sec:line_ratios}. 

\begin{figure}
	\centering 
	\hbox{
		\subfloat{\includegraphics[width = 0.5\textwidth, clip = true, trim = 0.0cm 0.15cm 0cm 0cm]{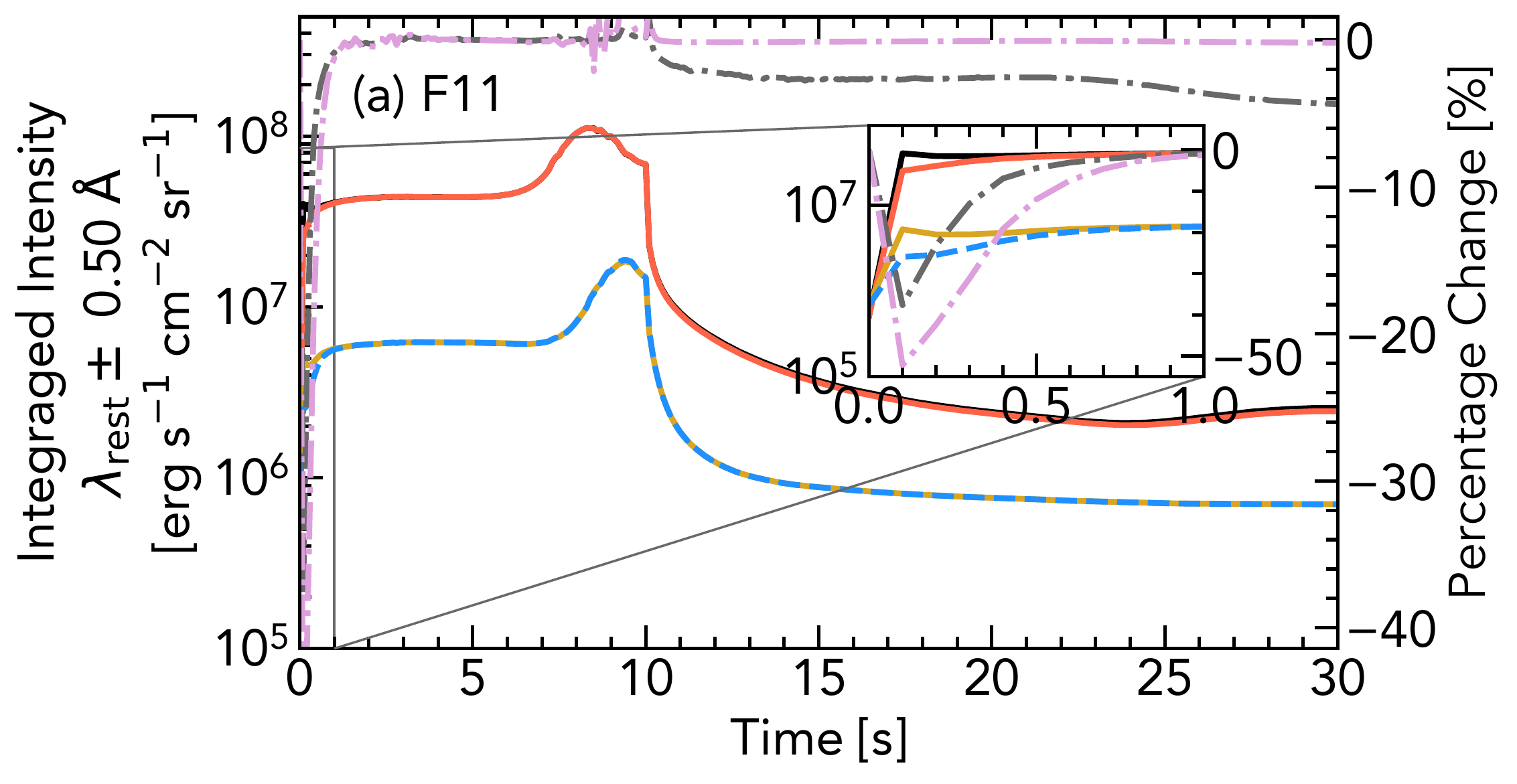}}	
		 }
	\hbox{
	        \subfloat{\includegraphics[width = 0.5\textwidth, clip = true, trim = 0.0cm 0.15cm 0cm 0cm]{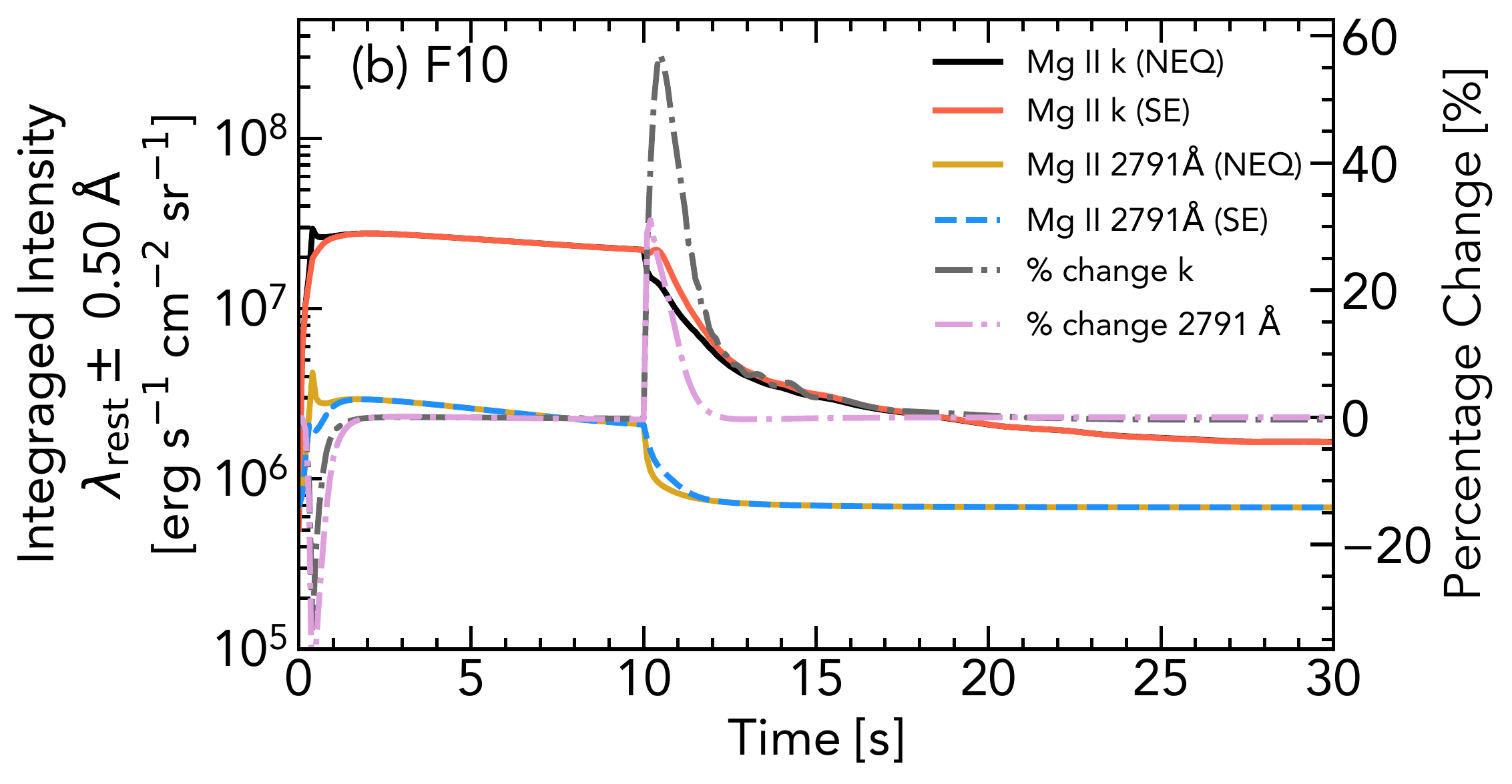}}	
                   }
        \hbox{
	        \subfloat{\includegraphics[width = 0.5\textwidth, clip = true, trim = 0.0cm 0.15cm 0cm 0cm]{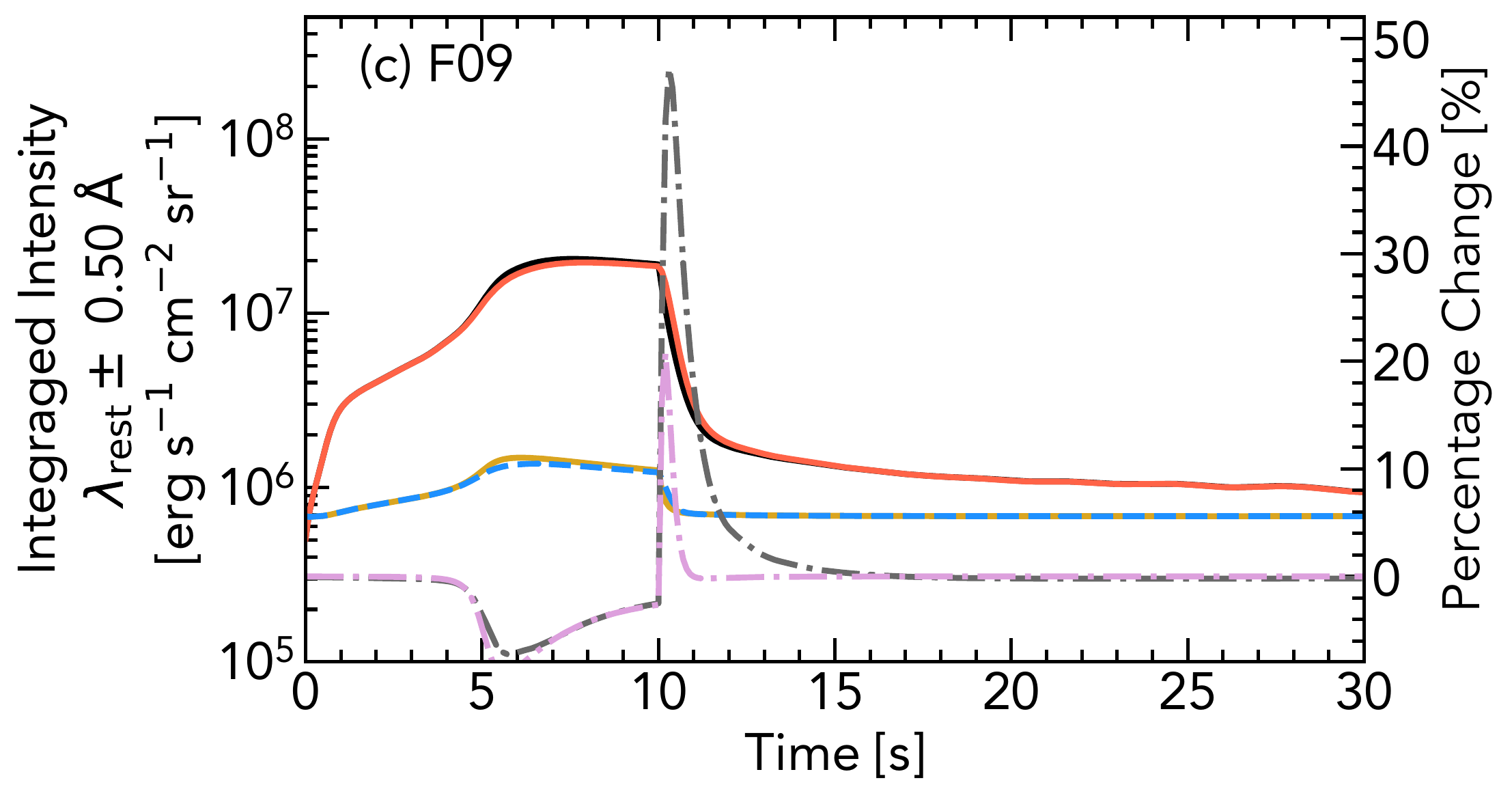}}	
                   }
	\caption{\textsl{Lightcurves of the Mg~\textsc{ii} k line $\pm0.5$~{\AA} in each simulation (a) F11, (b) F10, and (c) F09. The h line and subordinate lines behave qualitatively similar. An inset in panel (a) shows $t < 1$~s in more detail. In each panel: the black line is the k line NEQ solution, the red is the k line SE solution, the yellow line is the $2791$\AA\ NEQ solution, the blue dashed line is the 2791\AA\ SE solution, the grey dot-dashed line is the k line percentage change, and the purple dot-dashed line is the 2791\AA\ percentage change.}}
	\label{fig:lcurves}
\end{figure}

The largest differences appear around the emission peaks in the decay phase, with the wings showing less significant changes. Integrating across the line profiles therefore reduces the discrepancy at these times substantially. The discrepancies in the initial heating phase for the F10 and F11 simulations affected the line wings also, so the percentage change is of a similar magnitude to the specific intensity case at those times. Figures~\ref{fig:lcurves}(a,b,c) show the lightcurves of the Mg~\textsc{ii} k line, integrated $\pm 0.5$~\AA\ around the line core. 

In the moderate-to-strong flares the initial heating ($t<1$~s) shows a local maximum in the NEQ solution, but not in the SE solution. In the F10 flare, the NEQ solution shows a steeper decrease following the cessation of the beam. At other times the lightcurves generally show similar behaviour, with small intensity differences (note the logarithmic scales, so that while the lines do appear in close agreement they can actually differ by a few tens of percent).

%%%%%%%%%%%%%%%%%%%%%%%%%%%%%%%%%%%%%%%%%%%%%%%%%%%%%
%%%%%%%%%%%%%%%%%%%%     LINE RATIOS      %%%%%%%%%%%%%%%%%%%%%%%
%%%%%%%%%%%%%%%%%%%%%%%%%%%%%%%%%%%%%%%%%%%%%%%%%%%%%
\section{Line Ratios \& Formation Heights}\label{sec:line_ratios}
\begin{figure}
	\centering 
	{\includegraphics[width = 0.5\textwidth, clip = true, trim = 0.cm 0.cm 0.cm 0.cm]{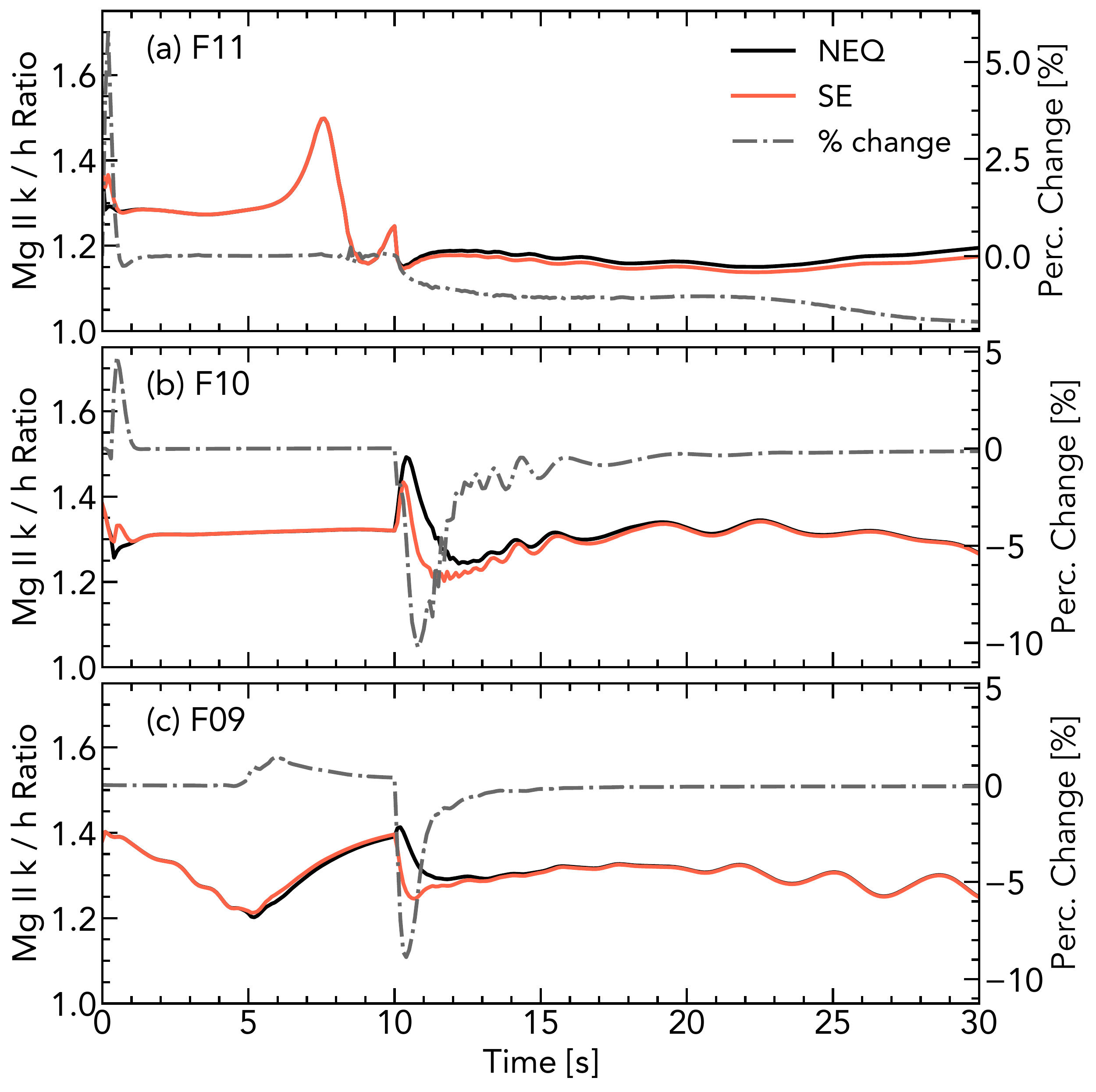}}	
	\caption{\textsl{Ratio of the k to h line ($R_{k:h}$) in each simulation. The red lines are the SE solution, black are the NEQ solution and grey dot-dashed lines shows the percentage change.}}
	\label{fig:ratios}
\end{figure}

The k:h line ratio, $R_{k:h}$, is a useful metric that can indicate if the lines are optically thick, if there are relative changes between the lines during the flare, or if radiative excitation processes are significant \citep[e.g.][]{2014ApJ...792...93H}. \cite{2015A&A...582A..50K} noted that in some areas of a flare ribbon observed by IRIS, $R_{k:h}$ changed in response to the flare. The ratio decreased slightly, and the relative spread in values decreased (there was a tighter correlation of $R_{k:h}$ than in areas outside of the ribbon). They speculated that this could be caused by the h \& k lines forming closer together than in the quiet Sun, and/or sampling a more uniform chromosphere during the flare. In the optically thin limit this ratio is $R_{k:h} = 2$, the ratio of the statistical weights of the k \& h upper levels. It is typical that $R_{k:h} \approx 1.2$, both in the quiet Sun and in flares \citep[e.g.][]{2015A&A...582A..50K,2015SoPh..290.3525L}, indicating optically thick line formation. Note, though, that $R_{k:h}$ can theoretically have a value of two even in the optically thick case as the ratio of the source functions can in effect take \textsl{any} value \citep{2015ApJ...811...80R}. 

It is important to determine if this ratio is affected by non-equilibrium effects that will confuse the interpretation of $R_{k:h}$ variation in flares if forward modelled using statistical equilibrium. 

Figure~\ref{fig:ratios} shows $R_{k:h}$ in each simulation where again we show the percentage change between NEQ and SE. The magnitude of the variations between the NEQ and SE solutions differs for the h \& k lines, meaning that $R_{k:h}$ consequently shows differences over time. The magnitude of this difference is relatively small, on the order of $<10$~\%. Generally the temporal profile of $R_{k:h}$ is preserved, though in the weaker flares the rate of change at the start of the decay phase is smaller in the NEQ solution, and in the F10 \& F11 simulations the NEQ ratios decrease somewhat whereas the SE ratios increase within the first second. Both NEQ and SE solutions have $R_{k:h}< 2$, indicating optically thick line formation, which we confirmed from inspecting the detailed line formation properties. 

\begin{figure}
	\centering 
	\hbox{
        \subfloat{\includegraphics[width = 0.5\textwidth, clip = true, trim = 0.cm 0.cm 0cm 0cm]{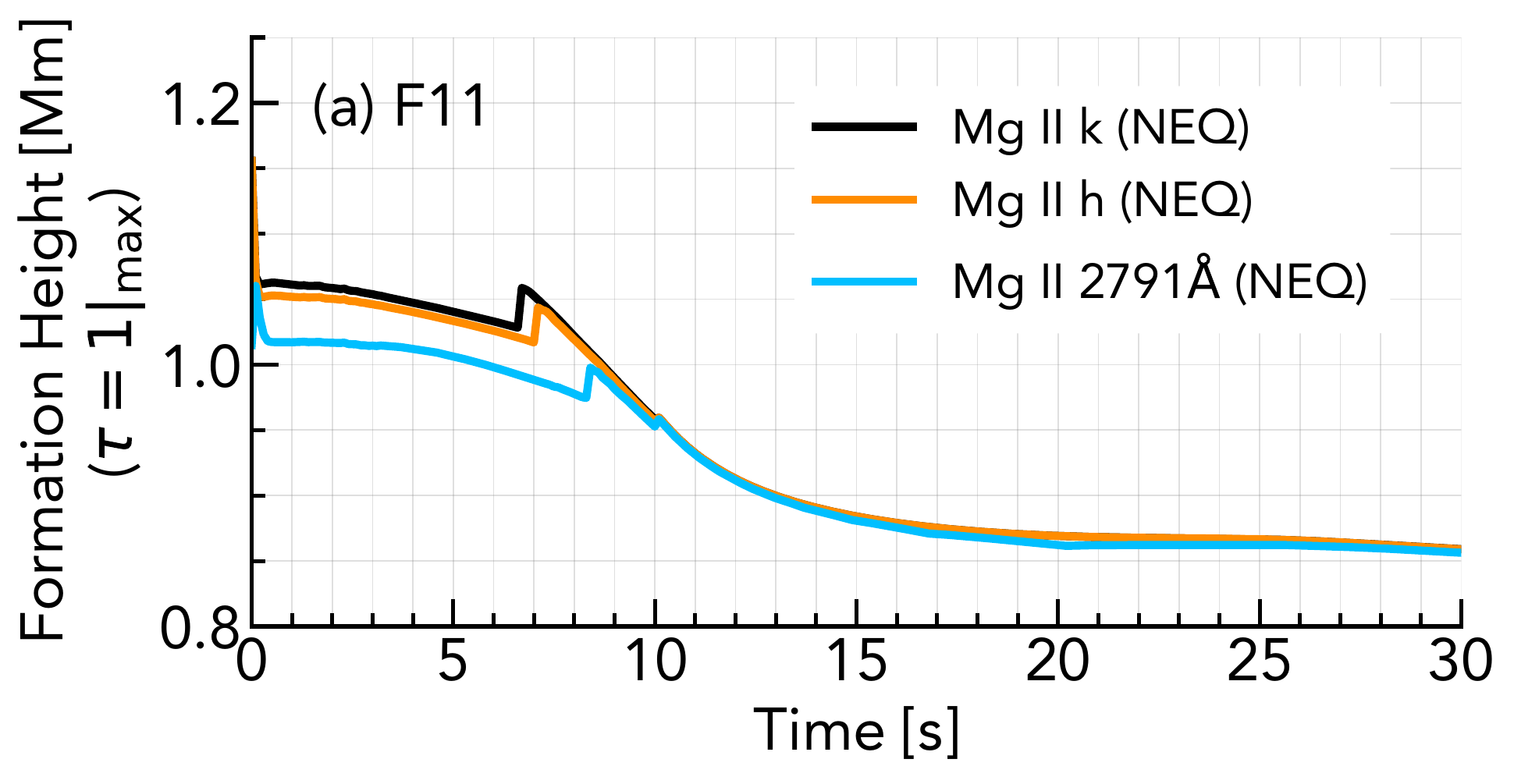}}
        }
        \hbox{
        \subfloat{\includegraphics[width = 0.5\textwidth, clip = true, trim = 0.cm 0.cm 0cm 0cm]{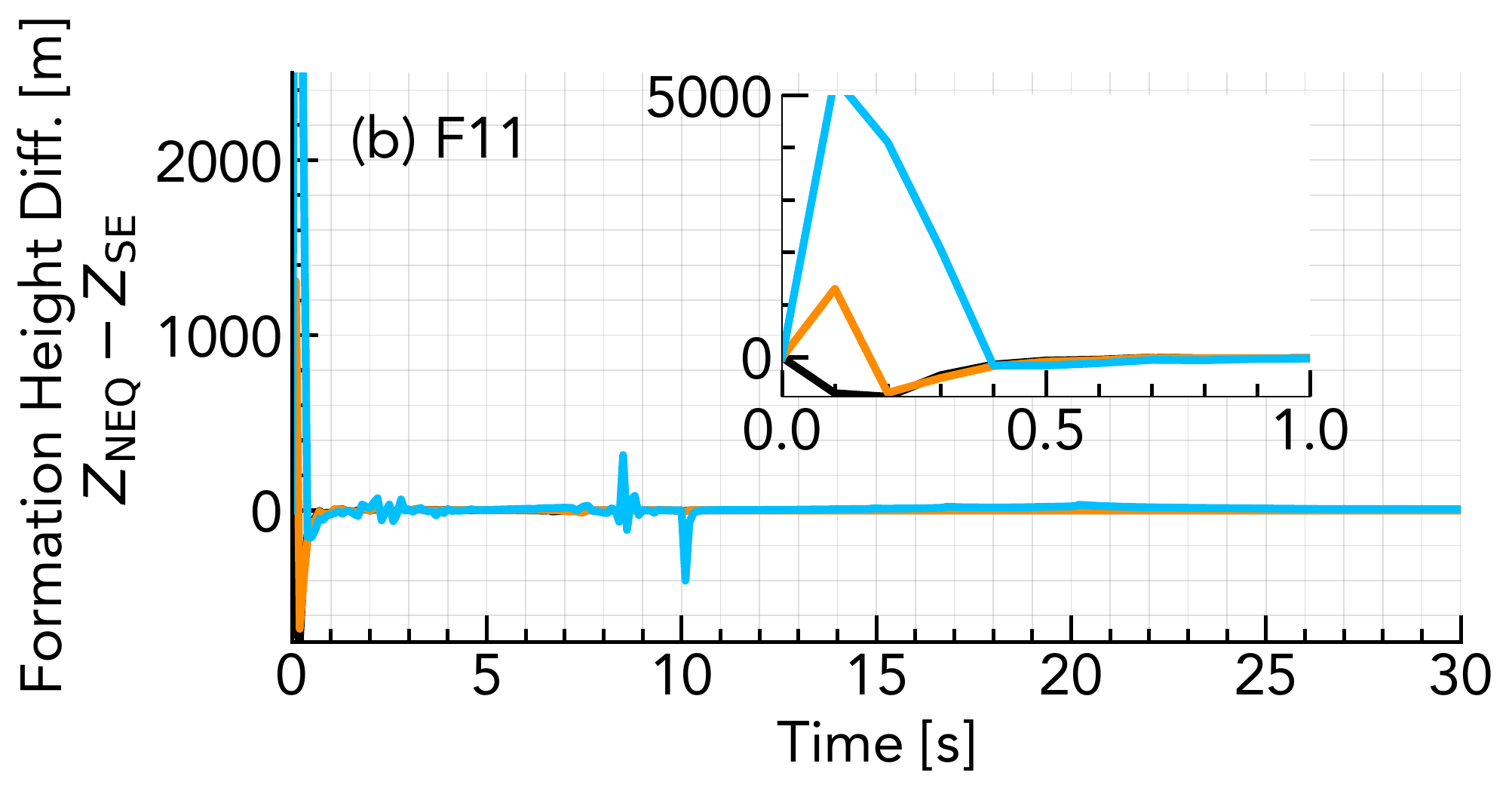}}
        }
        \hbox{
        \subfloat{\includegraphics[width = 0.5\textwidth, clip = true, trim = 0.cm 0.cm 0cm 0cm]{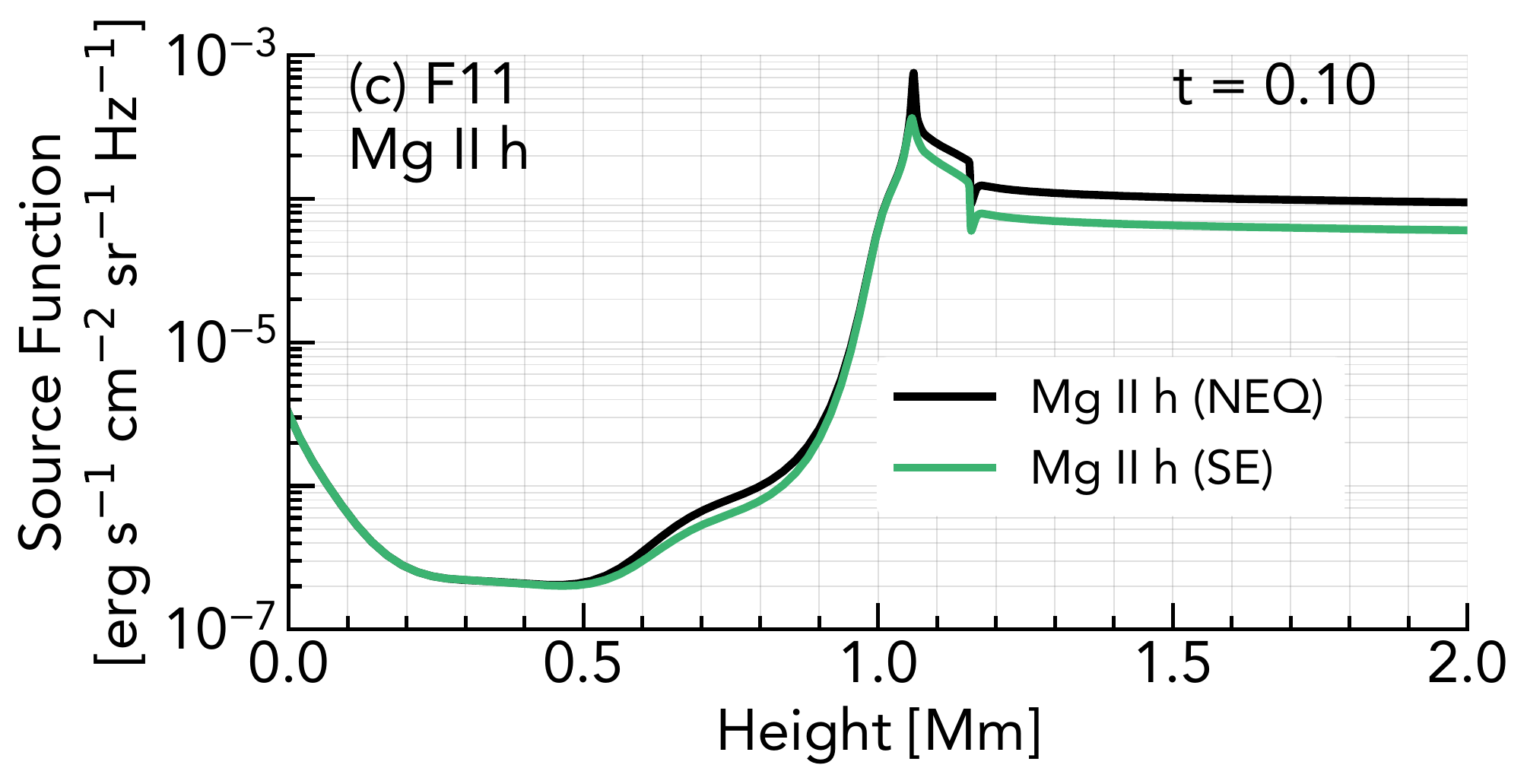}}	
        }
	\caption{\textsl{Panel (a) shows the formation height of the Mg~\textsc{ii}  h line (orange),  Mg~\textsc{ii} k line (black) and  Mg~\textsc{ii} 2791\AA\ line (blue) as a function of time in the F11 NEQ simulation. Panel (b) shows the formation height difference resulting from using the NEQ or SE solution, with the inset highlighting the first $t=0-1$~s of the simulation (note the change in scale to meters). Panel (c) shows the Mg~\textsc{ii} h line source function at $t = 0.1$~s, where black is the NEQ solution and green is the SE solution.}}
	\label{fig:f11lineform}
\end{figure}

\begin{figure}
	\centering 
	\hbox{
        \subfloat{\includegraphics[width = 0.5\textwidth, clip = true, trim = 0.cm 0.cm 0cm 0cm]{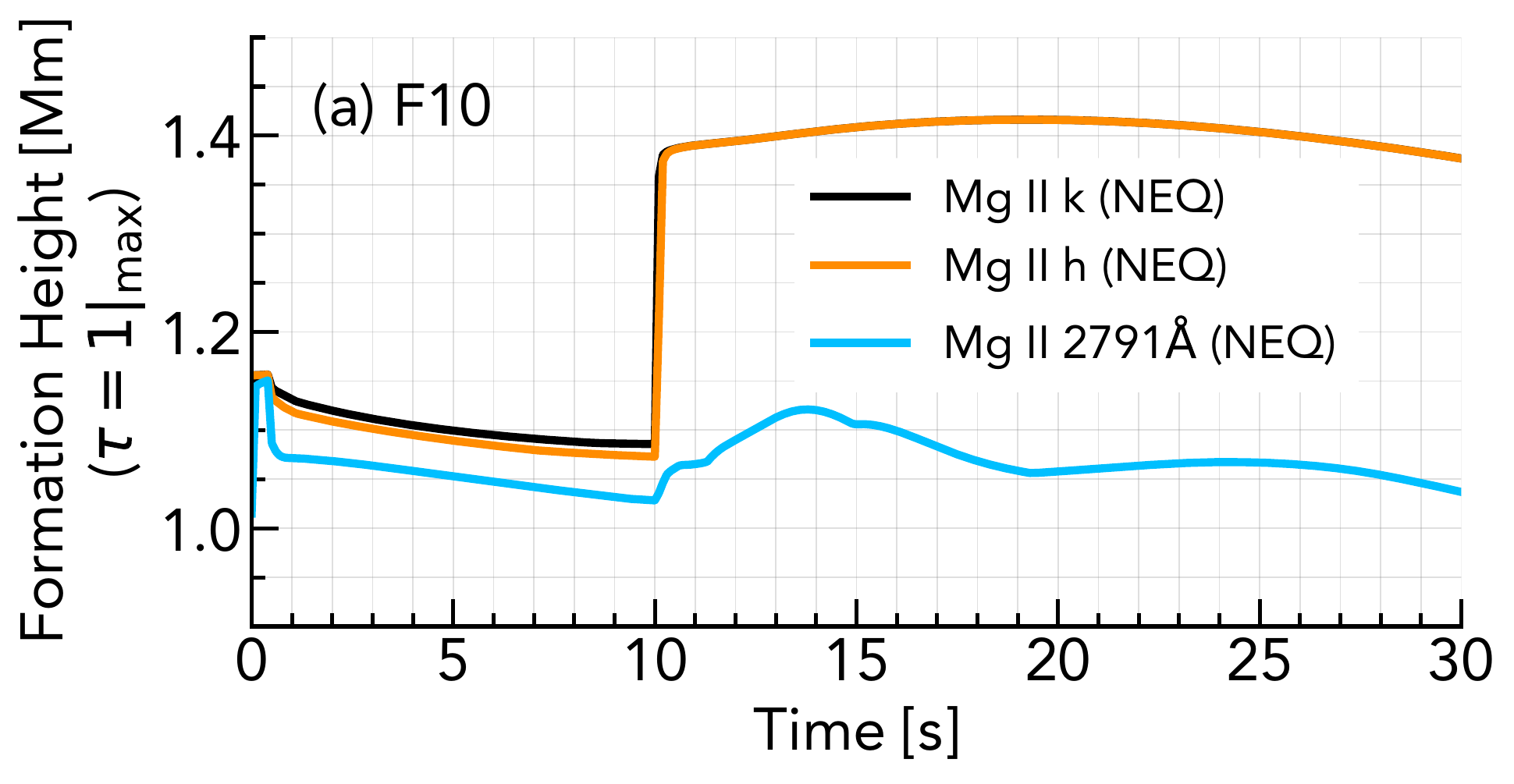}}
        }
        \hbox{
        \subfloat{\includegraphics[width = 0.5\textwidth, clip = true, trim = 0.cm 0.cm 0cm 0cm]{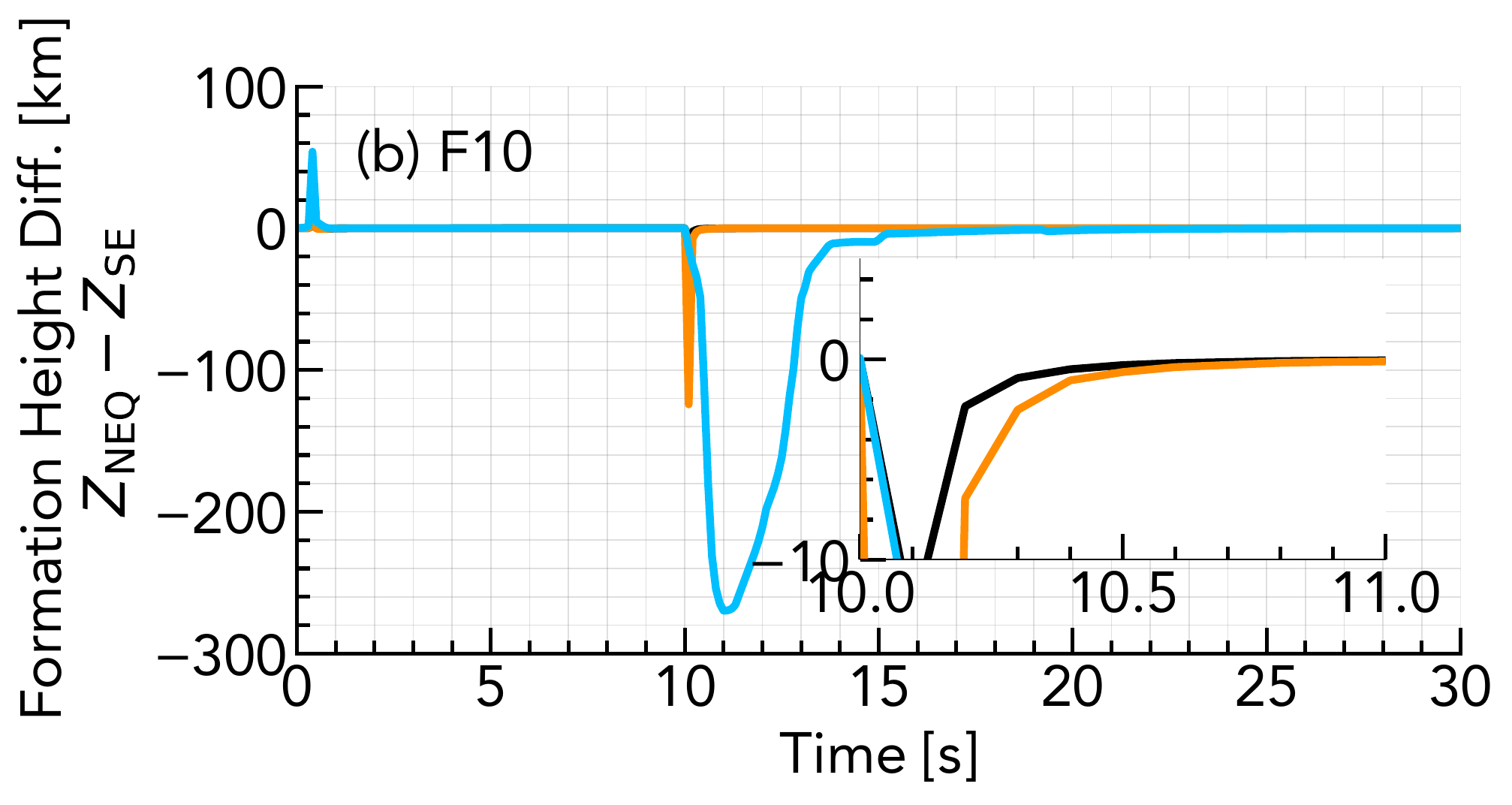}}
        }
        \hbox{
        \subfloat{\includegraphics[width = 0.5\textwidth, clip = true, trim = 0.cm 0.cm 0cm 0cm]{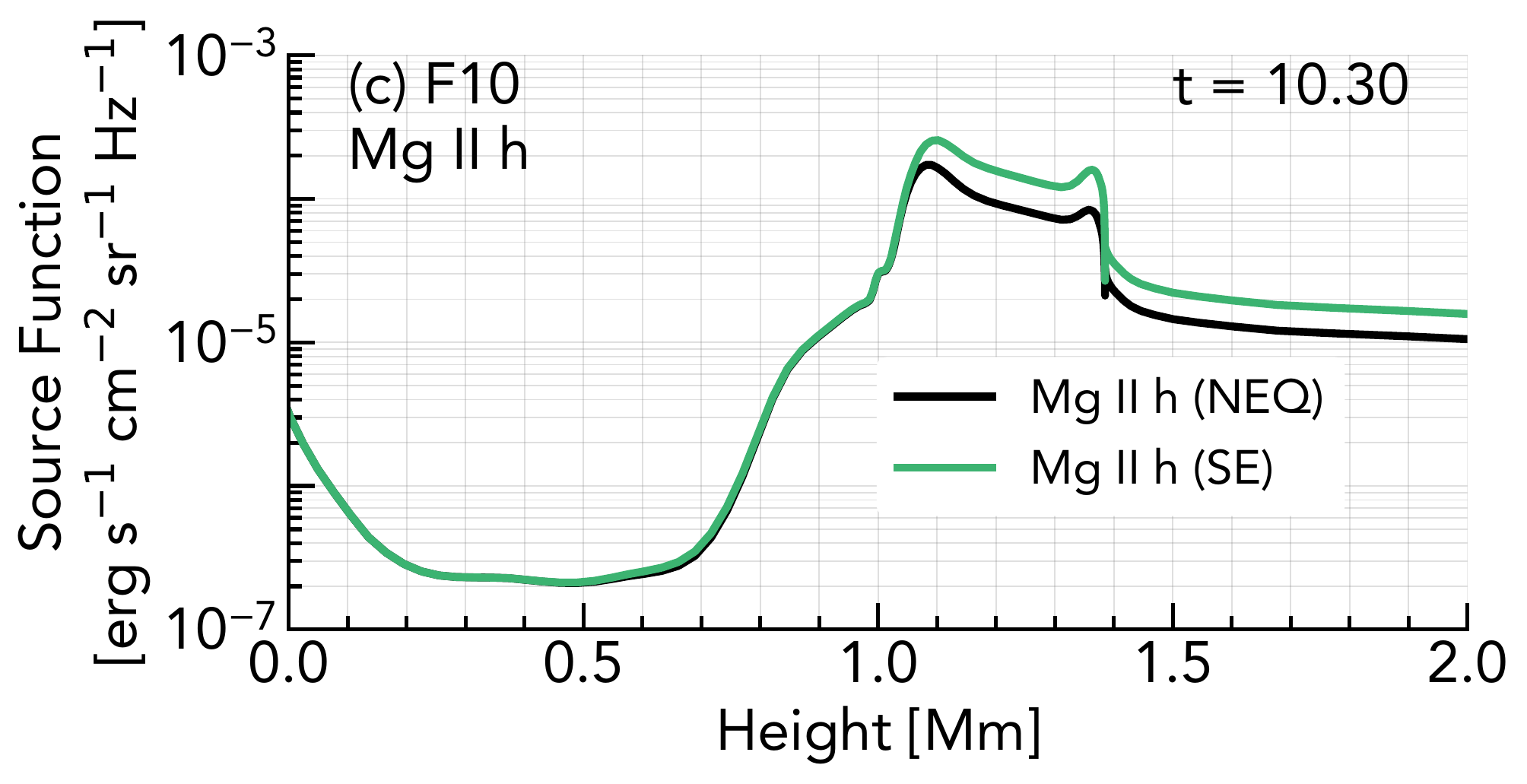}}	
        }
	\caption{\textsl{Same as Figure~\ref{fig:f11lineform}, but for the F10 simulation. Note also the change in scale in panel (b) compared to Figure~\ref{fig:f11lineform}.}}
	\label{fig:f10lineform}
\end{figure}

The ratio is influenced by the relative separation of the h \& k lines in the flare atmosphere, and the formation height of the lines is also useful in relating model results to observations of the flaring plasma. Figures~\ref{fig:f11lineform} \& ~\ref{fig:f10lineform} show  the formation heights of the h, k and 2791\AA\ lines in the NEQ solution (panels a), the formation height differences of the lines ($\Delta z$) between the NEQ and SE solutions (panels b), and the h line source function in the NEQ and SE solutions (panel c), for the F11 and F10 simulations respectively. Note that the scales vary between the two flares. The formation height here is defined as the height at which the $\tau_{\nu} = 1$ surface is maximal (that is, we are defining the line core to the part of the line forming highest in the atmosphere, with the greatest opacity). 

In both cases during the main phase of the flare there is little difference in the formation heights between the NEQ or SE solutions. However, at the very start of the heating phase in both flares the formation heights can differ. In the F11 simulation this is on the order of kilometres, and in the F10 simulation this is on the order tens to a few hundred km. The $\Delta z$ of k and h lines are of different magnitudes, and can be of different directions in the F11 simulation (the k line forms deeper in the NEQ solution whereas the h line forms higher, compared to the SE case). In both cases the subordinate line at 2791\AA\ actually shows a much greater formation height variation than the h \& k lines, forming a few hundred km deeper in the NEQ solution in the F10 flare during the initial decay phase. The h \& k lines in that flare do reach $\Delta z\sim100$~km, but this rapidly decreases to $\Delta z \sim5$~km or smaller.

Given the generally small differences in the ratios and formation heights, and their short lived nature, we do not envisage using SE will result in any significant misinterpretations when relating line profile features to plasma properties. A possible exception is the subordinate line during the decay phase in weak or moderate flare simulations, or if extremely strong gradients are present near the formation heights of the Mg~\textsc{ii} lines. 

%%%%%%%%%%%%%%%%%%%%%%%%%%%%%%%%%%%%%%%%%%%%%%%%%%%%%
%%%%%%%%%%%%%%%     LEVEL POPULATIONS     %%%%%%%%%%%%%%%%%%%%%%%
%%%%%%%%%%%%%%%%%%%%%%%%%%%%%%%%%%%%%%%%%%%%%%%%%%%%%
\section{Ion Fractions \& Relaxation Timescales}\label{sec:levelpops}
\begin{figure*}
	\centering 
	{\includegraphics[width = 0.9\textwidth, clip = true, trim = 0.cm 0.cm 0.cm 0.cm]{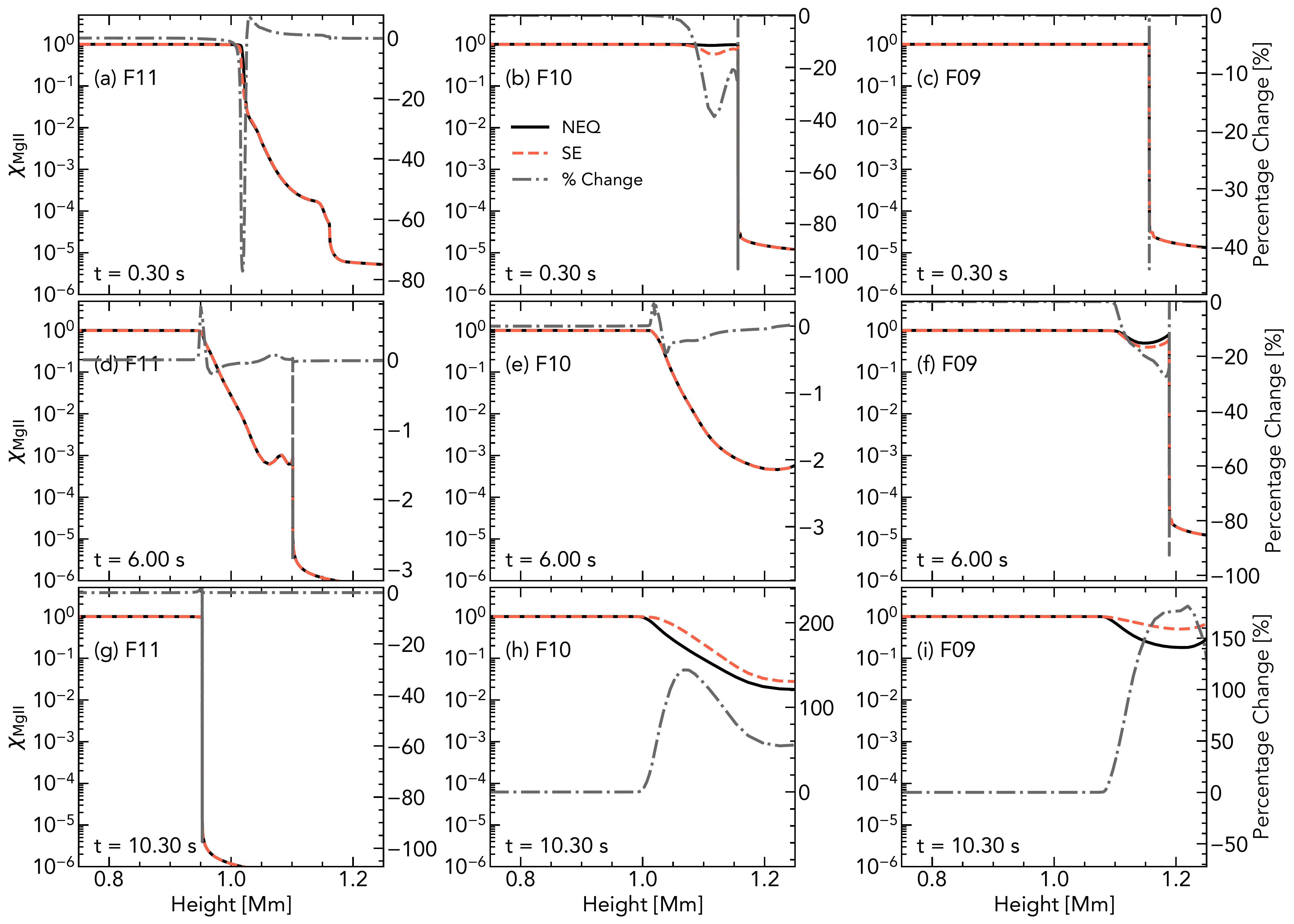}}	
	\caption{\textsl{Mg~\textsc{ii} ion fraction in each simulation at three different times. The top row is the initial heating phase ($t=0.3$~s), the middle row is the main heating phase ($t=6$~s), and the bottom row is the initial cooling phase ($t=10.3$~s). The black lines are the NEQ solutions, the red dashed line are the SE solutions, and the grey dot-dashed lines are the percentage change.}}
	\label{fig:ionfrac_comps}
\end{figure*}

The differences in the line profile results obtained from the NEQ and SE solutions discussed previously can be understood by studying the ion fraction stratification and the atomic level populations. 

Figures~\ref{fig:ionfrac_comps}(a,b,c) show the fraction Mg~\textsc{ii}/Mg, $\chi_{MgII}$, at the very start of the heating phase of the three simulations. In the two stronger flares, where the flare disturbs the atmosphere in a more impulsive and dramatic manner, $\chi_{MgII}$ is larger in the NEQ solution than the SE solution, and consequently there is a larger population in the resonance and subordinate line upper levels. This then results in more radiative decays, and more intense lines. In the F11 simulation there is a narrow region where there is up to an $80~\%$ change in $\chi_{MgII}$ between SE and NEQ. The discrepancy rapidly reduces, both in magnitude and spatial extent.  The F10 behaves in a similar manner. The weak flare, however, shows a difference only through the TR, slightly above the formation height of Mg~\textsc{ii}, so that the line intensity is not really affected.

During the main heating phase the NEQ and SE solutions give largely similar results with only small percentage changes in the F11 and F10 solutions.  Figures~\ref{fig:ionfrac_comps}(d.e,f) shows $t=6$~s, illustrating that in the main heating phase of the flares the NEQ and SE solutions have only marginal differences. In the F9 simulation, which evolved more slowly, differences have started to appear, explaining why there are line intensity differences at this stage in the flare. 

Finally, Figures~\ref{fig:ionfrac_comps}(g,h,i) shows the start of the cooling phase, shortly after the beam has ceased depositing energy, where the atmospheric temperature rapidly drops. Here the situation is reversed, with the SE solution predicting more $\chi_{MgII/Mg}$. The F11 simulation, in which the flare induced electron density and temperature enhancements were much larger, shows a smaller difference in comparison to the F10 and F9 simulations.

The ion fraction differences between the NEQ and SE solutions are due to the ionisation/recombination timescales in each simulation, which vary with atmospheric state (temperature and electron density).

We determined the timescales for the ionisation equilibrium of Mg~\textsc{ii}/Mg~\textsc{iii} following the methodology of \cite{2002ApJ...572..626C} and \cite{2013ApJ...772...89L}: For a selected time step of a flare simulation, the temperature was increased by $1\%$ throughout the atmosphere and the rate equations for the full magnesium model atom were solved as a function of time (keeping the hydrodynamic state constant at the perturbed value), following the relaxation of the population densities from the initial state towards the new equilibrium. The relaxation time scale was calculated from a fit of the time evolution of the population density of Mg~\textsc{iii} to the analytic solution for a two-level atom:

\begin{equation}\label{eq:relaxtime}
n(t)=n(\infty)+(n(0)-n(\infty))e^{-t/\tau_{\mathrm{relax}}},
\end{equation}

\noindent where $n(t)$ is the population density of Mg~\textsc{iii} at time t, $n(\infty)$ is the statistical equilibrium population density in the perturbed atmosphere, $n(0)$ is the equilibrium population density in the initial atmosphere, and $\tau_{\mathrm{relax}}$ is the relaxation timescale.

\begin{figure*}
	\centering 
	{\includegraphics[width = 0.75\textwidth, clip = true, trim = 0.cm 0.cm 0.cm 0.cm]{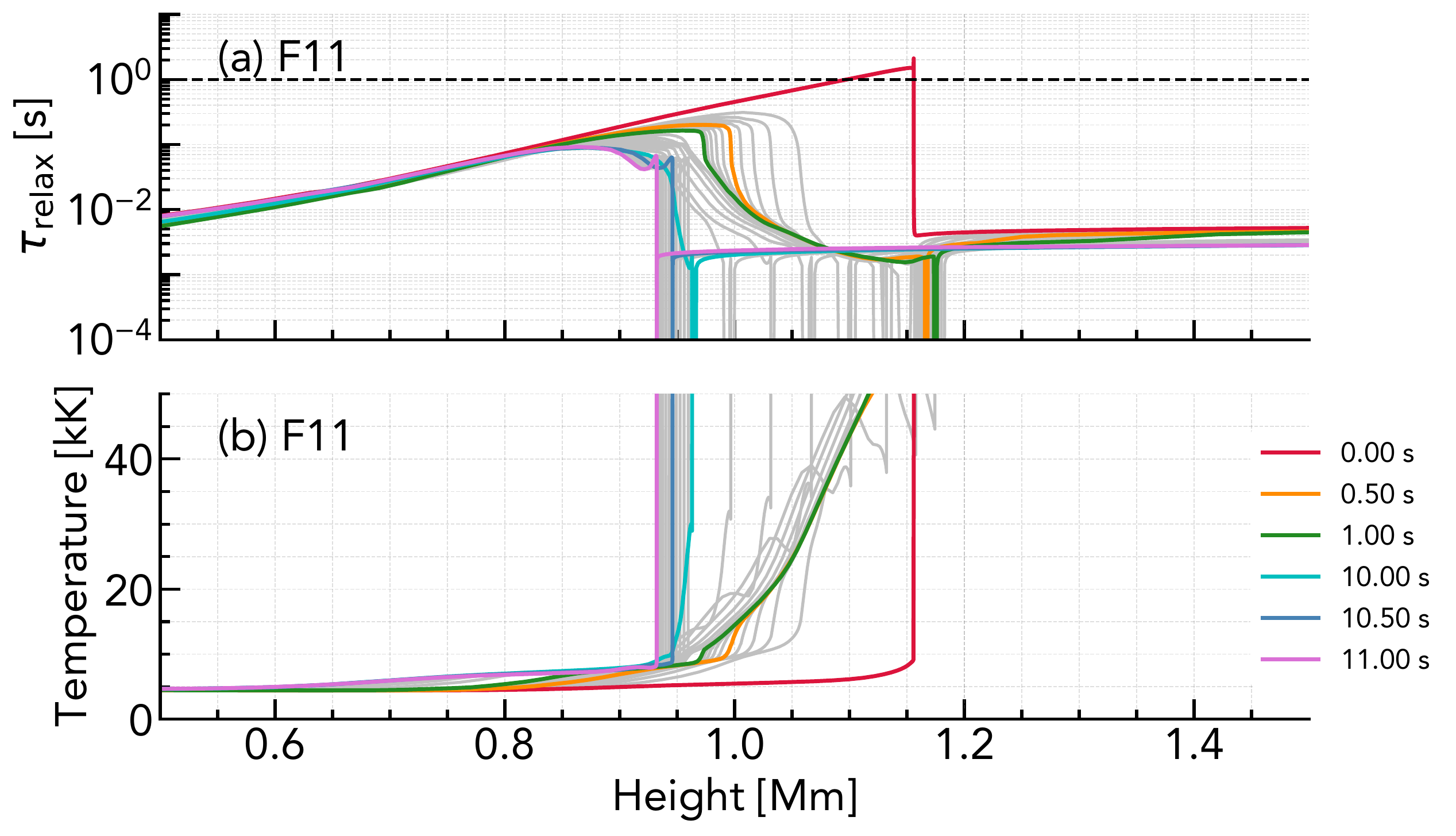}}	
	\caption{\textsl{\textsl{In each panel the relaxation timescale (a) and temperature (b) are shown at various snapshots in the F11 simulation. Several times are highlighted by coloured lines, as indicated in the legend.}}}
	\label{fig:f11_timescale}
\end{figure*}
\begin{figure*}
	\centering 
	{\includegraphics[width = 0.75\textwidth, clip = true, trim = 0.cm 0.cm 0.cm 0.cm]{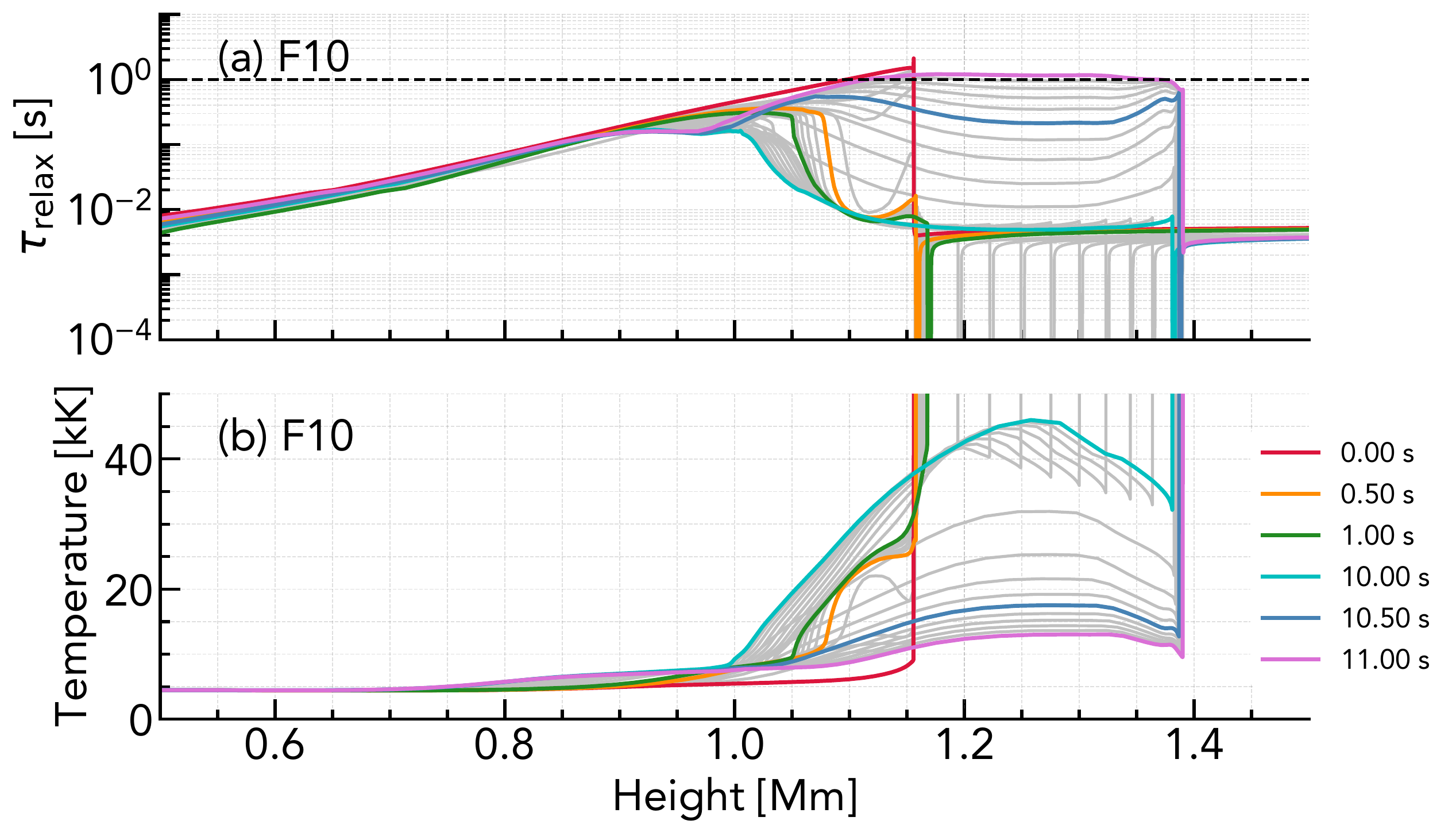}}	
	\caption{\textsl{\textsl{Same as Figure~\ref{fig:f11_timescale}, but for the F10 simulation.}}}
	\label{fig:f10_timescale}
\end{figure*}
\begin{figure*}
	\centering 
	{\includegraphics[width = 0.75\textwidth, clip = true, trim = 0.cm 0.cm 0.cm 0.cm]{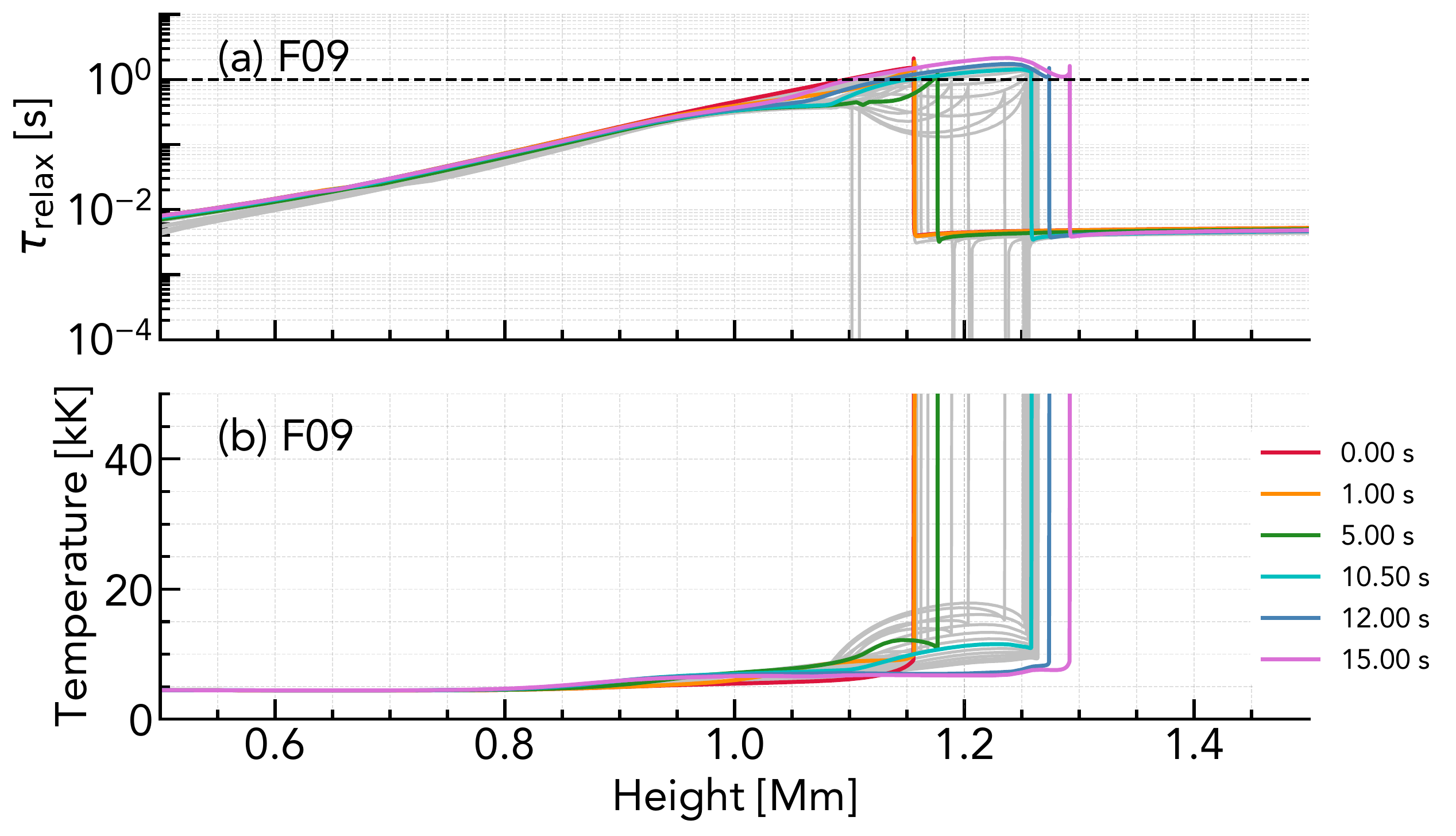}}	
	\caption{\textsl{\textsl{Same as Figure~\ref{fig:f11_timescale}, but for the F09 simulation.}}}
	\label{fig:f09_timescale}
\end{figure*}

Figures~\ref{fig:f11_timescale},~\ref{fig:f10_timescale}, \& ~\ref{fig:f09_timescale} show the relaxation timescales (top panels) and temperatures (bottom panels) at various times in the three simulations. The relaxation timescales are initially on the order of $\tau_{\mathrm{relax}} \sim 1-2$~s in the upper chromosphere. Through the first second of flare heating in the F11 and F10 simulations, the relaxation time decreases to the order of $\tau_{\mathrm{relax}} < 0.1$s in response to the substantial temperature and electron density increase in the chromosphere. NEQ effects are prominent within this initial second of heating, since even though the relaxation timescale is decreasing, the atmosphere evolves very impulsively. In the main heating phase the atmosphere evolves somewhat more slowly, so that Mg~\textsc{ii} ionisation equilibrium keeps pace given the small  $\tau_{\mathrm{relax}}$.

In the F9 simulation  $\tau_{\mathrm{relax}} = 1$ does not decrease as much as in the stronger flares, but the atmosphere also evolves more gradually so that NEQ effects only become apparent when the dynamics become faster. At these times  $0.1 < \tau_{\mathrm{relax}} < 1$~s, and Mg~\textsc{ii} is somewhat out of equilibrium.

During the decay phase of the F9 and F10 simulations  $\tau_{\mathrm{relax}}$ rises back to the order of 1s when the temperature drops, so that there is initially a large disagreement between NEQ and SE while the atmosphere is cools rapidly. After the initial sharp decrease in temperature (and electron density), the rate of change of the atmosphere where Mg II forms is smaller so that the differences between NEQ and SE get smaller (but are still present to some degree), and Mg~\textsc{ii} is only partially out of equilibrium.
 
In the F11 case the temperature is so much larger at the end of the heating phase that even through the decay phase the relaxation timescale is still sufficiently small that the discrepancies are reduced.

%%%%%%%%%%%%%%%%%%%%%%%%%%%%%%%%%%%%%%%%%%%%%%%%%%%%%
%%%%%%%%%%%%%%%     NEQ POPS->RH      %%%%%%%%%%%%%%%%%%%%%%%%%%
%%%%%%%%%%%%%%%%%%%%%%%%%%%%%%%%%%%%%%%%%%%%%%%%%%%%%

%%%%%%%%%%%%%%%%%%%%%%%%%%%%%%%%%%%%%%%%%%%%%%%%%%%%%
%%%%%%%%%%%     SUMMARY & CONCLUSIONS      %%%%%%%%%%%%%%%%%%%%%%%
%%%%%%%%%%%%%%%%%%%%%%%%%%%%%%%%%%%%%%%%%%%%%%%%%%%%%
\section{Summary \& Conclusions}\label{sec:conclusions}
Non-equilibrium effects on the formation of Mg~\textsc{ii} spectra during solar flares of different magnitudes has been investigated using \texttt{RADYN} and \texttt{MS\_RADYN} simulations. The time-dependent NLTE NEQ atomic level populations and synthetic spectra were computed, and compared to the NLTE SE solution.

While line profile shapes are preserved in each solution, the intensities of lines can differ, sometimes substantially, if NEQ effects are taken into consideration. These largely appear in the initial heating phase ($t<1$~s), and in the initial seconds of the decay phase. Investigation of Mg~\textsc{ii} ionisation equilibrium showed that changes in the atmospheric state can lower the relaxation timescale, meaning that Mg~\textsc{ii} can be very close to equilibrium during the main heating phase, but that when the temperature rapidly falls Mg~\textsc{ii} is again driven out of equilibrium. 

In \cite{kerr_2019} we found that partial frequency redistribution is required to accurately forward model Mg~\textsc{ii} in flares, with intensity changes of $200-1000$~\% between the CRD and PRD solutions, and certain features only appearing due to redistribution. With \textsc{MS\_RADYN} we are able to include NEQ effects, but the line profiles are computed assuming CRD.  PRD is seemingly more important to include than NEQ effects. While the magnitudes of differences between NEQ and SE was generally smaller than the differences between CRD and PRD, and shorter lived (PRD was required throughout the duration of the flare), if one is interested in the initial flare heating, or in the decay phase, then NEQ effects should ideally be considered. 

A time dependent code capable of including both NEQ and PRD effects in flares should be developed, but this is a computationally demanding endeavour. For the moment, if we wish to study the initial heating and cooling, we can capture both NEQ and PRD by using a multi-step process. Radiative hydrodynamic flare atmospheres are produced by \texttt{RADYN}. The \texttt{RADYN} solutions are used in \texttt{MS$\_$RADYN} to obtain the NEQ CRD Mg~\textsc{ii} populations. Those NEQ populations are fed into \texttt{RH} which, using the modifications we described in \cite{kerr_2019} to fix the level populations, will solve the PRD Mg~\textsc{ii} radiation transfer.  
	
Another useful flare line observed by IRIS is Fe \textsc{xxi} 1354.1~\AA, which forms in 10MK plasma and is likely to experience non-equilibrium effects \citep{2011ApJS..194...26B}. It is commonplace to model this line using data from the \texttt{CHIANTI} atomic database \citep{1997A&AS..125..149D,2019ApJS..241...22D}, under the assumption of ionisation equilibrium and optically thin emission \citep[e.g.][]{2015ApJ...799..218Y,2019ApJ...879L..17P}. While the latter assumption is safe, ignoring non-equilibrium effects may not be correct. A similar comparison to that presented here for Mg~\textsc{ii} should be performed, comparing the predicted level Fe~\textsc{xxi} ion fraction from \texttt{MS\_RADYN} to those from \texttt{CHIANTI} given model atmospheres from \texttt{RADYN}. \\

\textsc{Acknowledgments:} \small{GSK was funded by an appointment to the NASA Postdoctoral Program at Goddard Space Flight Center, administered by USRA through a contract with NASA. This research was supported by the Research Council of Norway through its Centres of Excellence scheme, project number 262622, and through grants of computing time from the Programme for Supercomputing. JCA acknowledges funding support from the Heliophysics Supporting Research and Heliophysics Innovation Fund programs.}

\bibliographystyle{aasjournal}
\bibliography{Kerr_etal_nonequilMgII}

\end{document}